\documentclass[12pt,letterpaper,a4paper]{article}
\usepackage[includeheadfoot,
            marginratio={1:1,2:3},
            width=460pt,
            height=700pt,]{geometry}
\usepackage{amsmath}
\usepackage{amsfonts}
\usepackage{amssymb}
\usepackage{graphicx}
\usepackage{cancel}
\usepackage{empheq}
\usepackage{color}
\usepackage{hyperref}

\usepackage{float}
\restylefloat{table}
\restylefloat{figure}
\usepackage[utf8]{inputenc}
\usepackage{amsmath, amsthm, amssymb, amsfonts}
\usepackage[font=small, labelfont=bf]{caption}
\usepackage{amsmath, amsthm, amssymb, amsfonts}
\usepackage{multirow}
\usepackage{graphicx}
\usepackage{float}
\usepackage[numbers,sort&compress]{natbib}
\usepackage{enumerate}
\usepackage{amsmath}
\usepackage{amsfonts}
\usepackage{amssymb}
\usepackage{graphicx}
\usepackage{cancel}
\usepackage{empheq}
\usepackage{color}
\usepackage{hyperref}
\usepackage{float}
\restylefloat{table}
\restylefloat{figure}

\numberwithin{equation}{section}

\newcommand{\nc}{\newcommand}
\nc{\beq}{\begin{equation}}
\nc{\eeq}{\end{equation}}
\nc{\bea}{\begin{eqnarray}}
\nc{\eea}{\end{eqnarray}}

\def\ov{\overline}

\begin{document}
{\hfill
IFT-UAM/CSIC-17-121

\hfill 
arXiv:1712.07310}

\vspace{1.0cm}
\begin{center}
{\Large
Symplectic formulation of the type IIA \vskip0.2cm nongeometric scalar potential}
\vspace{0.4cm}
\end{center}

\vspace{0.35cm}
\begin{center}
 Xin Gao$^\dag$\footnote{Email: xgao@roma2.infn.it}, Pramod Shukla$^\ddag$\footnote{Email: pramod.shukla@uam.es}, Rui Sun$^\diamond$\footnote{Email: rsun@tsinghua.edu.cn}
\end{center}

\vspace{0.1cm}
\begin{center}
{
$^{\dag}$Dipartimento di Fisica, Universita di Roma ``Tor Vergata", and \\
I.N.F.N. Sezione di Roma ``Tor Vergata",\\
Via della Ricerca Scientifica, 00133 Roma, Italy \\
\vskip0.4cm
$^{\ddag}$Departamento de F\'{\i}sica Te\'orica and Instituto de F\'{\i}sica Te\'orica UAM/CSIC,\\
Universidad Aut\'onoma de Madrid, Cantoblanco, 28049 Madrid, Spain\\
\vskip0.4cm
$^{\diamond}$Yau Mathematical Sciences Center, Tsinghua University, Haidian District, \\Beijing 100084, China
}
\end{center}

\vspace{1cm}
\begin{abstract}
We study the four-dimensional (4D) scalar potential arising from a generalized type IIA flux superpotential including the (non-)geometric fluxes. First, we show that using a set of peculiar flux combinations, the 4D scalar potential can be formulated into a very compact form. This is what we call as the `symplectic formulation' from which one could easily anticipate the ten-dimensional origin of the effective scalar potential. We support our formulation through an alternate derivation of the scalar potential via considering the Double Field Theory (DFT) reduction on a generic Calabi Yau orientifold. In addition, we also exemplify the insights of our formulation with explicit computations for two concrete toroidal examples using orientifolds of the complex threefolds ${\mathbb T}^6/{({\mathbb Z}_2 \times {\mathbb Z}_2)}$ and ${\mathbb T}^6/{\mathbb Z}_4$.
\end{abstract}

\clearpage

\tableofcontents

\section{Introduction}
\label{sec_intro}
Type II supergravity theories admit generalized fluxes via a successive application of T-duality on the three-form $H_3$ which results in a chain of geometric and non-geometric fluxes given as under \cite{Hellerman:2002ax,Dabholkar:2002sy,Hull:2004in,Derendinger:2004jn, Derendinger:2005ph, Shelton:2005cf, Wecht:2007wu},
\bea
\label{eq:Tdual}
& & H_{ijk} \longrightarrow \omega_{ij}{}^k  \longrightarrow Q_{i}{}^{jk}  \longrightarrow R^{ijk}\, .
\eea
Generically, such fluxes can appear as parameters in the four-dimensional (4D) effective potential, and subsequently can help in developing a suitable scalar potential which could be useful for various model building purposes. Moreover, interesting connections among the toolkits of superstring flux-compactifications, the gauged supergravities and the Double Field Theory (DFT) via non-geometric fluxes have given the platform for approaching phenomenology based goals from these three directions  \cite{ Derendinger:2004jn, Derendinger:2005ph, Shelton:2005cf, Wecht:2007wu, Aldazabal:2006up, Dall'Agata:2009gv, Aldazabal:2011yz, Aldazabal:2011nj, Geissbuhler:2011mx, Grana:2012rr, Dibitetto:2012rk, Andriot:2013xca, Andriot:2014qla, Blair:2014zba, Andriot:2012an, Geissbuhler:2013uka}.

A consistent incorporation of the various possible fluxes makes the compactification background richer and more flexible for model building. In this regard, a continuous progress has been made since more than a decade towards moduli stabilization \cite{Aldazabal:2006up, Ihl:2006pp, Ihl:2007ah, Blumenhagen:2015kja}, in constructing de-Sitter vacua \cite{deCarlos:2009qm, Danielsson:2012by, Blaback:2013ht, Damian:2013dwa, Blumenhagen:2015xpa} and also in realizing minimal aspects of inflationary cosmology \cite{Damian:2013dq, Hassler:2014mla, Blumenhagen:2014gta, Blumenhagen:2015qda}. It is surprisingly remarkable that almost all the phenomenological studies have been made using non-geometric 4D effective potentials via merely knowing the forms of the K\"ahler- and the super-potentials, and without having a full understanding of their ten-dimensional origin. However, in the meantime some significant steps have also been taken in parallel towards exploring the form of non-geometric ten-dimensional action via DFT \cite{Andriot:2013xca, Andriot:2011uh, Andriot:2012wx} as well as supergravity \cite{Villadoro:2005cu, Blumenhagen:2013hva, Gao:2015nra, Shukla:2015bca}. In this regard, toroidal orientifolds have been always in the center of attraction because of their relatively simpler structure to perform explicit computations, and so toroidal setups have served as promising toolkits; e.g. a ten-dimensional origin of the four dimensional  scalar potential with geometric flux $(\omega)$ in type IIA toroidal orientifold has been invoked in \cite{Villadoro:2005cu}. This has been subsequently generalized with the inclusion of non-geometric $(Q, R)$-fluxes for type IIA- and non-geometric $(Q)$-flux for IIB-theory in \cite{Blumenhagen:2013hva} and the resulting oxidized 10D action was found to be compatible with  DFT action. 

There have been close connections between the symplectic geometry and effective potentials of type II supergravity theories \cite{Ceresole:1995ca, D'Auria:2007ay}, and the role of symplectic geometry gets crucially important while dealing with Calabi Yau (CY) orientifolds. The reason for the same being the fact that unlike toroidal orientifold examples, one does not know the explicit analytic representation of the Calabi Yau metric needed to express the effective potential. For example, in the context of type IIB orientifolds with the presence of standard NS-NS three-form flux ($H_3$) and RR three-form flux ($F_3$), the two scalar potentials, one arising from the  $F/D$-term contributions while the other being derived from the dimensional reduction of the ten-dimensional kinetic pieces, could be matched via merely using the period matrices and without the need of knowing CY metric \cite{Taylor:1999ii, Blumenhagen:2003vr}. Similarly an extensive study of the effective actions in symplectic formulation have been made for both type IIA as well as type IIB superstring compactifications in the presence of standard fluxes using Calabi Yau threefolds and their orientifolds \cite{Grimm:2004ua, Grimm:2004uq, Benmachiche:2006df}. 

Unlike the studies of \cite{Villadoro:2005cu, Blumenhagen:2013hva, Gao:2015nra, Shukla:2015bca} which need the knowledge of explicit internal background metric and hence hard to concretely promote to the Calabi Yau case, in the context of type IIB non-geometric flux compactification, the symplectic formulation of \cite{Taylor:1999ii, Blumenhagen:2003vr} has been recently extended with the inclusion of various (non-)geometric fluxes in \cite{Shukla:2015hpa} and also the so-called non-geometric S-dual `P-flux' \cite{Shukla:2016hyy}. In the meantime, a very robust analysis has been performed by considering the DFT reduction on generic Calabi Yau threefold, and subsequently the generic ${\cal N} =2$ results have been used to derive the ${\cal N} = 1$ type IIB effective potential with non-geometric fluxes \cite{Blumenhagen:2015lta}. In all these studies it has been found that the $F$-term and $D$-term contributions of the 4D effective scalar potential combines in a specific way such that they correspond to some kinetic pieces of a ten-dimensional action.

\subsection*{Motivation and main goals}
Most of the studies regarding exploring the non-geometric ten-dimensional uplift of the 4D effective potential have been performed for type IIB orientifold compactifications. For the ten-dimensional non-geometric action of type IIA case, all the known formulations are written out using internal background metric, e.g. see \cite{Villadoro:2005cu, Blumenhagen:2013hva}. In this article, we plan to study along the following lines:
\begin{itemize}
\item{First, we plan to present a symplectic formulation for the 4D non-geometric type IIA effective potential obtained from a generalized flux superpotential and some $D$-terms, for a generic Calabi Yau orientifolds compactifications. Subsequently we dimensionally oxidize the scalar potential to invoke its ten-dimensional origin. This compact symplectic formulation could also be helpful in performing a model independent moduli stabilization.}
\item{We implement the robust ${\cal N} =2$ results of DFT reduction on Calabi Yau threefolds \cite{Blumenhagen:2015lta} to provide an alternate derivation of our symplectic formulation ensuring that there is indeed a higher dimensional theory which upon dimensional reduction results in the same non-geometric type IIA scalar potential derived from a generalized flux superpotential.}
\item{We demonstrate the insights of our proposal via two concrete threefold ($X_3$) examples; one with considering the orientifold of a ${\mathbb T}^6/{({\mathbb Z}_2 \times {\mathbb Z}_2)}$ complex threefold while the other with that of a ${\mathbb T}^6/{\mathbb Z}_4$ complex threefold. The latter was considered to illustrate the $D$-terms as the former setup is too simple to support the $D$-term which needs a non-trivial even sector in the $(1,1)$-cohomology, i.e. $h^{1,1}_+(X_3) \neq 0$ \cite{Robbins:2007yv}.}
\end{itemize}
\noindent
The article is organized as follows: In section \ref{sec_basics} we provide the relevant preliminaries for type IIA non-geometric flux compactifications. In section \ref{sec_NeworbitsAndscalarpotential}, we present a set of peculiar flux combinations which we subsequently use for presenting a symplectic formulation of the scalar potential, and its dimensional oxidation. Section \ref{sec_twoExs} presents two concrete examples to demonstrate the specific insights of our proposal using the orientifolds of two toroidal complex threefolds, namely ${\mathbb T}^6/{({\mathbb Z}_2 \times {\mathbb Z}_2)}$ and ${\mathbb T}^6/{\mathbb Z}_4$. In section \ref{sec_DFT} we present a DFT derivation of our symplectic formulation of the effective 4D scalar potential. Finally the important conclusions and outlook are presented in section \ref{sec_conclusions}. 

\section{Non-geometric Type IIA setup: Preliminaries}
\label{sec_basics}
In this work, we consider type IIA superstring theory compactified on an orientifold of a Calabi Yau threefold $X_3$ with the presence of $O6$-planes. In this regard, the orientifold is constructed via modding out the CY with a
discrete symmetry ${\cal O}$ which includes the world-sheet parity $\Omega_p$ combined with the space-time fermion number in the left-moving sector $(-1)^{F_L}$. In addition
${\cal O}$ can act non-trivially on the Calabi-Yau manifold so that one has altogether,
\bea
\label{eq:orientifoldO}
& & {\cal O} = \Omega_p\, (-1)^{F_L}\, \sigma
\eea
where $\sigma$ is an involutive symmetry (i.e. $\sigma^2=1$) of the internal CY and acts trivially on the four flat dimensions.
The massless states in the four dimensional effective theory are in one-to-one correspondence with various involutively even/odd harmonic forms, and hence do generate the equivariant cohomology groups $H^{(p,q)}_\pm(X_3)$. Subsequently, the various field ingredients can be expanded in appropriate bases of the equivariant cohomologies. For that purpose, let us first start with fixing the conventions. 

\subsection{Fixing the conventions}
Let us mention that we will be following the notations of \cite{Ihl:2007ah, Robbins:2007yv} with some small symbolic changes due to the same being utilized for multiple purposes. To begin with, we consider the following representations for the various involutively even and odd harmonic forms \cite{Grimm:2004ua},
\begin{table}[H]
\begin{center}
\begin{tabular}{|c|| c| c| c| c| c| c|} 
\hline
&&&&&&\\
Cohomology group & $H^{(1,1)}_+$ & $H^{(1,1)}_-$ & $H^{(2,2)}_+$ & $H^{(2,2)}_-$ & $H^{(3)}_+$ & $H^{(3)}_-$ \\
&&&&&&\\
\hline\hline
&&&&&&\\
 Dimension & $h^{1,1}_+$ & $h^{1,1}_-$ & $h^{1,1}_-$ & $h^{1,1}_+$ & $h^{2,1}+ 1$ & $h^{2,1} + 1$ \\
 &&&&&&\\
 Basis & $\mu_\alpha$ & $\nu_a$ & $\tilde\nu^a$ & $\tilde\mu^\alpha$ & $\alpha_I$ & $\beta^J$ \\
 &&&&&&\\
 \hline
\end{tabular}
\end{center}
\caption{Representation of various forms and their counting}
\label{tab_1}
\end{table}
\noindent
Here $\mu_\alpha$ and $\nu_a$ denote the bases of even and odd real harmonic two-forms respectively, while $\tilde\mu^\alpha$ and $\tilde\nu^a$ denote the bases of  odd and even four-forms. Further, $\alpha_I$ and $\beta^J$ form the bases of even and odd real three-forms. In addition, the zero form {\bf 1} is even while there is an involutively odd six-form $\Phi_6$. Moreover, we consider the following intersection among the basis of various forms,
\bea
\label{eq:intersectionBases}
& & \int_{X_3} \Phi_6 = f, \qquad \int_{X_3} \nu_a \wedge \nu_b \wedge \nu_c= \kappa_{abc} , \qquad \int_{X_3} \nu_a \wedge \mu_\alpha \wedge \mu_\beta = \hat{\kappa}_{a \alpha \beta},\\
& & \qquad \int_{X_3} \nu_a \wedge \tilde{\nu}^b= d_a{}^b, \qquad \int_{X_3} \mu_\alpha \wedge \tilde{\mu}^\beta = \hat{d}_\alpha{}^\beta , \qquad \int_{X_3} \alpha_I \wedge \beta^J=\delta_I{}^J\,. \nonumber
\eea
Note that our notations are a bit more generic and flexible for appropriate normalization of forms. Of course, for an appropriate choice of the bases of four-forms to be dual to the respective two-forms, one would have $d_a{}^b = \delta_a{}^b$ and $\hat{d}_\alpha{}^\beta = \delta_\alpha{}^\beta$.

In order to preserve ${\cal N} =1$ supersymmetry, one needs the involution $\sigma$ to be anti-holomorphic, isometric and acting on the K\"ahler form $J$ as under,
\bea
\label{eq:sigmaJ}
\sigma^\ast(J) = - J\,,
\eea
which generically results in the presence of $O6$-planes. Given that the K\"ahler form $J$ and the NS-NS two-form potential $B_2$ are odd under the involution, the same can be expanded in the odd two-form basis $\nu_a$ as,
\bea
& & J = t^a \, \nu_a\, , \qquad \qquad B_2 = b^a\, \nu_a\,.
\eea
Similarly, the nowhere vanishing holomorphic three-form ($\Omega_3$) of the Calabi Yau threefold can be expanded in the three-form basis as under,
\bea
& & \Omega_3 = {\cal X}^K\, \alpha_K - {\cal F}_K\, \beta^K\, .
\eea
Now, the compatibility of $\sigma$ with the Calabi Yau condition $(J \wedge J \wedge J) \propto \, (\Omega_3 \wedge \ov \Omega_3)$ demands the following condition,
\bea
\label{eq:sigmaOmega}
\sigma^\ast(\Omega_3) =  e^{2\, i \theta} \, \ov \Omega_3\, \, \, ,
\eea
which implies that
\bea
\label{eq:OrientifoldOmega}
& & {\rm Im}(e^{-i\,\theta}\, {\cal X}^K) = 0, \qquad {\rm Re}(e^{-i\,\theta}\, {\cal F}_K) = 0 \,.
\eea
In addition, note that only one of these equations is relevant due to the scale invariance of $\Omega_3$ which is defined only up to a complex rescaling. In order to take care of this, we fix our convention by choosing the followings,
\bea
\label{eq:Omega=1}
& & \int_{X_3} i\, \Omega_3 \wedge \ov \Omega_3 = 1, \qquad \sigma^\ast(\Omega_3) = \ov \Omega_3,
\eea
which simply means setting $\theta$ in eqn. (\ref{eq:sigmaOmega}) to zero implying that ${\cal X}^K$ are real functions of the complex structure moduli while ${\cal F}_K$'s are pure imaginary, and altogether satisfy 
\bea
& & {\cal X}^K {\cal F}_K = -i/2\, .
\eea 
Now we can define the following two complexified variables ($T^a$ and $N^K$) using the complexified K\"ahler form $J_c$ and complexified three-form $\Omega_c$ which are given as,
\bea
& & J_c \equiv B_2 + i\, J = T^a\, \nu_a\,,
\eea
and
\bea
& & \Omega_c \equiv C_3 + 2\, i \, e^{-D} {\rm Re}(\Omega_3) = \xi^K\, \alpha_K + 2\, i \, e^{-D}\, {\cal X}^K\, \alpha_K \, = 2\, N^K\, \alpha_K \,.
\eea
Here $D$ is the four-dimensional dilaton which can be given in terms of ten-dimensional dilaton $\hat\phi$ by the relation,
\bea
& & e^{2 \, D} = \frac{e^{2 \hat\phi}}{{\cal V}}\,.
\eea
In addition, we have used that the RR three-form potential $C_3$ contains a piece $\xi^K \, \alpha_K$ in its expansion.

\subsection{Four-dimensional ${\cal N}=1$ scalar potential}
The bosonic part of the effective action for a ${\cal N} = 1$ supergravity theory having one gravity multiplet, a set of complex scalars $\varphi^{\cal A}$ and a set of vectors $A^\alpha$ can be given as \cite{Grimm:2004ua},
\bea
& & S^{(4)} = - \int_{M_4} \left( -\frac{1}{2} \, R \ast 1 + K_{{\cal A} \ov{{\cal B}}} \, d\varphi^{\cal A} \wedge \ast d{\ov{\varphi}}^{\ov {\cal B}} + V \ast 1 \right) \nonumber\\
& & \hskip2cm + \frac{1}{2} \left({\rm Re} {f_g} \right)_{\alpha\beta} \, F^\alpha \wedge \ast F^\beta + \frac{1}{2} \left({\rm Im} {f_g} \right)_{\alpha\beta} \, F^\alpha \wedge F^\beta \,,
\eea
where $\ast$ is the four-dimensional Hodge star, and $F^\alpha = d A^\alpha$. Further, the total scalar potential can be given as a sum of $F$-term and $D$-term contributions as under:
\bea
\label{eq:Vtotal}
V \equiv V_F + V_D \, , 
\eea
where
\bea
\label{eq:VF}
& & V_F =  e^K \left(K^{{\cal I}\ov{\cal I}^\prime}\, D_{\cal I} W \, \ov{D}_{\ov{\cal I}^\prime} \ov{W} - 3 \, |W|^2\right)\,,
\eea
and
\bea
\label{eq:VD}
& & V_D = \frac{1}{2} \, \left( {\rm Re} f_g \right)^{\alpha\beta}\, D_\alpha \, D_\beta \,.
\eea
Here the sum is over all the moduli and the covariant derivative is defined to be $D_{\cal I} = d_{\cal I} + W\, \partial_{\cal I} K$, and $D_\alpha$ is the $D$-term for the $U(1)$ gauge group corresponding to $A^\alpha$ given as under,
\bea
& & D_\alpha = \left(\partial_{{\cal A}} K\right) \, {\left({\cal T}_\alpha\right)}^{\cal A}{}_{\cal B} \, \varphi^{\cal B} + \zeta_\alpha \,,
\eea
where ${\cal T}_\alpha$ is the generator of the gauge group and $\zeta_\alpha$ denotes the Fayet-Iliopoulos term. For the present work we will split the index ${\cal I}$ as ${\cal I} = \{I, a\}$ where the capital letters $I$ are counted in $\{0, 1, .., h^{2,1}(X_3) \}$ and the small letters $a$'s are counted by $h^{1,1}_-(X_3)$.

Now, as the four dimensional scalar potential is determined by three main ingredients, namely the K\"ahler potential ($K$), the superpotential ($W$) and the holomorphic gauge kinetic function ($f_g$), let us elaborate more on these ingredients.
\subsubsection*{The K\"ahler potential ($K$)}
The K\"ahler potential consists of two pieces which are given as \cite{Grimm:2004ua},
\bea
\label{eq:Kgen}
& & K = 4\, D - \ln\left(8 \, {\cal V}\right) \, ,
\eea
where ${\cal V} = \frac{1}{6} \kappa_{abc}\, t^a \,t^b \, t^c$. Thus the K\"ahler potential can be thought of as a real function of the complexified moduli $T^a$ and $N^K$ implicitly appearing through $t^a$ and the dilaton $D$.

\subsubsection*{The flux superpotential ($W$)}
For getting the generalized version of GVW flux superpotential \cite{Gukov:1999ya}, we need to define a twisted differential operator ${\cal D}$ as under,
\bea
\label{eq:twistedD}
& & {\cal D} = d + H \wedge . - \omega \triangleleft . +Q \triangleright . - R \bullet .
\eea
Here the action of various fluxes appearing in ${\cal D}$ is such that for an arbitrary $p$-form $A_p$, the pieces $H\wedge A_p$, $\omega \triangleleft A_p$, $Q \triangleright A_p$ and $R \bullet A_p$ denote a $(p+3)$-form, a $(p+1)$-form, a $(p-1)$-form and a $(p-3)$-form respectively. More specifically, there are the following flux actions on various harmonic-forms \cite{Ihl:2007ah},
\bea
\label{eq:fluxActions0}
& & \hskip-1.3cm H \wedge {\bf 1} = H_K\, \beta^K, \qquad \qquad \qquad \qquad \, \, \, H \wedge \alpha_K = - f^{-1}\, H_K \,  \Phi_6 \nonumber\\
& &  \hskip-1.3cm \omega \triangleleft \nu_a = \omega_{a K}\, \beta^K, \qquad \qquad \qquad \qquad \, \, \, \omega \triangleleft \mu_\alpha = \hat{\omega}_{\alpha}^K\, \alpha_K, \nonumber\\
& &  \hskip-1.4cm \omega \triangleleft \alpha_K = {(d^{-1})}_a{}^b\, \omega_{b K} \, \tilde\nu^a, \qquad \qquad \quad \, \, \, \omega \triangleleft \beta^K = -{(\hat{d}^{-1})}_\alpha{}^\beta\, \hat\omega_{\beta}^K \, \tilde\mu^\alpha \,,\nonumber\\
& & \\
& & \hskip-1.3cm Q \triangleright \tilde\nu^a = Q^{a}_{K}\, \beta^K, \qquad \qquad \qquad \qquad \, \, Q \triangleright  \tilde\mu^\alpha = \hat{Q}^{\alpha K}\, \alpha_K, \nonumber\\
& & \hskip-1.3cm Q \triangleright \alpha_K = -{(d^{-1})}_a{}^b\, Q^{a}_{K} \, \nu_b, \qquad \qquad \, \, Q \triangleright \beta^K = {(\hat{d}^{-1})}_\alpha{}^\beta\, \hat{Q}^{\alpha K} \, \mu_\beta\,, \nonumber\\
& & \hskip-1.3cm R \bullet {\Phi_6} = R_K\, \beta^K, \qquad \qquad \qquad \qquad R \bullet \alpha_K =  f^{-1}\, R_K \,  {\bf 1} \,, \nonumber
\eea
where we also note that $H\wedge \beta^K = 0 = R \bullet \beta^K$. In addition, the usual RR and NS-NS fluxes can be expanded as under,
\bea
\label{eq:flux-components}
& & \hskip-1cm F_0 = m_0, \quad F_2 = m^a\, \nu_a, \quad F_4 = e_a\, \tilde{\nu}^a, \quad F_6 = e_0\, \Phi_6 \,; \quad   H_3 = H_K \, \beta^K\,.
\eea
Now, the generic tree level superpotential has two pieces given as under,
\bea
\label{eq:Wrrnsns}
& & \hskip1.5cm W = \,W_{RR} + W_{NS-NS}\,,\\
& & \hskip-3.6cm {\rm where} \nonumber\\
& & \hskip -0.6cm W_{RR}=\int_{X_3} e^{J_c} \wedge F_{RR} , \qquad W_{NS-NS} = \int_{X_3} \Omega_c \wedge {\cal D}\left( e^{- J_c} \right)\,.\nonumber
\eea
Now we consider the followings
\bea
& & e^{J_c} = 1 + J_c + \frac{1}{2} J_c \wedge J_c + \frac{1}{3!} \, J_c \wedge J_c \wedge J_c \, ,
\eea
and
\bea
& & F_{RR} = F_0 + F_2 + F_4 + F_6 \,.
\eea
Utilizing the flux actions of various NS-NS and RR fluxes on various cohomology bases as given in eqns. (\ref{eq:flux-components}) and (\ref{eq:fluxActions0}), the superpotential takes the following form with explicit dependence on the complexified moduli \cite{Grimm:2004ua, Ihl:2007ah},
\bea
\label{eq:Wgen}
&  W_{RR} &= f\, \, e_0 + d_a{}^b \, T^a \, \, e_b + \frac{1}{2}\, \kappa_{abc} \, T^a \,T^b\, \, m^c  + \frac{1}{6}\, \kappa_{abc}\, T^a \, T^b\, T^c\, \, m_0 \, , \\
&  W_{NS-NS} &= 2 N^K \biggl[ H_K + \omega_{aK} T^a + \frac{1}{2} \kappa_{abc} T^b T^c \left({(d^{-1})}_d{}^a\, Q^d{}_K \right) + \frac{1}{6}\, \kappa_{abc} T^a T^b T^c  \left(f^{-1} R_K \right) \biggr] \,,\nonumber
\eea
where all the complexified moduli and axions are encoded in the following definitions of complexified variable $T^a$ and $N^K$,
\bea
\label{eq:chiralvariables}
& & T^a = b^a + i\, t^a\, , \qquad \qquad \qquad N^K = \frac{\xi^K}{2}\,+ i \, e^{-\phi}\, \sqrt{\cal V}\, \, {\cal X}^K\,.
\eea
Utilizing the generic form of the K\"ahler potential (\ref{eq:Kgen}) and the superpotential (\ref{eq:Wgen}), the $F$-term contribution to the four dimensional scalar potential is determined by eqn. (\ref{eq:VF}) where the sum is over all the $T^a$ and $N^K$ moduli. 

\subsubsection*{The gauge kinetic function ($f_g$) and $D$-term effects}
The $D$-term contribution to the scalar potential is determined by the gauge kinetic couplings. The same has two (so-called) electric and magnetic components given as,
\bea
\label{eq:fg}
& & (f_g^{\rm ele})_{\alpha\beta} = i\, \hat{\kappa}_{a\alpha\beta} \, T^a, \qquad (f_g^{\rm mag} )^{\alpha\beta}= -\, i \, \hat{\kappa}^{\delta\gamma} \hat{d}_{\delta}{}^\alpha \, \hat{d}_\gamma{}^\beta \, ,
\eea
where $\hat{\kappa}^{\delta\gamma}$ is the inverse of the shorthand notation $\hat{\kappa}_{\delta\gamma} = \hat{\kappa}_{\delta\gamma a}\, T^a$. Keeping in mind that four-dimensional vectors descend from reduction on the three-form $C_3$ while the dual four-form gauge fields arise from reduction on five-form potential $C_5$,  let us consider the following expansions of the $C_3$ and the $C_5$ \cite{Robbins:2007yv, Blumenhagen:2015lta},
\bea
& & C_3 = \xi^K \, \alpha_K + A^\alpha\, \mu_\alpha, \qquad C_5 =  A_\alpha\, \tilde\mu^\alpha \, .
\eea
Further we consider a pair $(\lambda^\alpha, \lambda_\alpha)$ to ensure the 4D gauge transformations of the quantities $(A^\alpha, A_\alpha)$ as under,
\bea
\label{eq:gaugeA}
& & A^\alpha  \to A^\alpha + d \lambda^\alpha, \qquad A_\alpha  \to A_\alpha + d \lambda_\alpha\,.
\eea
Subsequently considering the twisted differential ${\cal D}$ given in eqn. (\ref{eq:twistedD}), we find the following transformation of the RR forms,
\bea
\label{eq:C3change}
& & \hskip-0.9cm C_{RR} \equiv C_1 + C_3 + C_5 = \xi^K \, \alpha_K + A^\alpha\, \mu_\alpha + A_\alpha\, \tilde\mu^\alpha \nonumber\\
& & \longrightarrow C_{RR} + {\cal D}\left(\lambda^\alpha \, \mu_\alpha + \lambda_\alpha \, \tilde\mu^{\alpha} \right) \\
& & = \left(\xi^K - \lambda^\alpha \, \hat\omega_{\alpha}{}^K + \lambda_\alpha\, \hat{Q}^{\alpha K}\right) \, \alpha_K + A^\alpha\, \mu_\alpha + A_\alpha\, \tilde\mu^\alpha \, , \nonumber
\eea
where we have used the flux actions given in eqn. (\ref{eq:fluxActions0}). 
Now the transformation given in eqn. (\ref{eq:C3change}) shows that the scalar field $\xi^K$ is not invariant under the gauge transformation, and leads to the following shift in the ${\cal N} = 1$ coordinate $N^K$,
\bea
& & \delta N^K = -\frac{1}{2} \, \lambda^\alpha\, \hat\omega_{\alpha}{}^K + \frac{1}{2} \, \lambda_\alpha \, \hat{Q}^{\alpha K}\,.
\eea
This, in particular, implies that if we define a field as $\Xi_K = e^{i N^K}$,
then $\Xi_K$ is electrically charged under the gauge group $U(1)_\alpha$ with charge $-\frac{1}{2} \, \hat\omega_{\alpha}^K$ while it is magnetically charged with charge $\frac{1}{2} \, \hat{Q}^{\alpha K}$. Now using the relation $\left(\partial_K D\right) = - \, e^D \,{\cal F}_K$, this results into the following two D-terms \cite{Ihl:2007ah} (see \cite{Shukla:2015bca, Blumenhagen:2015lta} also for $D$-terms in type IIB orientifolds),
\bea
& & D_\alpha =  \frac{i}{2} \, \left(\partial_K K\right)\, \hat\omega_\alpha{}^K = -2 \, i \, e^D \, {\cal F}_K \, \hat\omega_\alpha{}^K, \\
& & \hskip0cm  D^\alpha = \frac{i}{2} \, \left(\partial_K K\right)\, \hat{Q}^{\alpha K} = 2 \, i \, e^D \, {\cal F}_K \, \hat{Q}^{\alpha K} \,. \nonumber
\eea
Subsequently, using these two $D$-terms and the respective gauge kinetic couplings in eqn. (\ref{eq:fg}), the scalar potential given in eqn. (\ref{eq:VD}) can be generically given as \cite{Ihl:2007ah},
\bea
\label{eq:DtermGen}
& & \hskip-1.5cm V_D = - 2\, e^{2\, D} \, {\cal F}_J \, {\cal F}_K \, \biggl(\left[{\rm Re}(f_g^{\rm ele})_{\alpha\beta} \right]^{-1} \, \hat{\omega}_\alpha{}^K \, \hat{\omega}_\beta{}^J + \left[{\rm Re}{(f_g^{\rm mag})}^{\alpha\beta}\right]^{-1} \, \hat{Q}^{\alpha K} \, \hat{Q}^{\beta J} \biggr) \,.
\eea
Note that ${\rm Re}(f_g^{\rm ele}) > 0$, ${\rm Re}(f_g^{\rm ele}) >0$ and ${\cal F}_I$'s are pure imaginary, and therefore $V_D \geq 0$.

The moduli dynamics of the four dimensional effective theory is determined by the total scalar potential given as a sum of $F$- and $D$-term contributions,
\bea
V_{tot} = V_F + V_D \,,
\eea
subject to satisfying some NS-NS Bianchi identities and a set of RR tadpole cancellation conditions.

\subsubsection*{Constraints from tadpoles cancellations and Bianchi identities}
Generically, there are tadpole terms present due to presence of $O6$-planes, and these can be canceled either by imposing a set of flux constraints or else by adding the counter terms which could arise from the presence of local sources such as (stacks of) $D6$-brane. These effects equivalently provide the following contributions in the effective potential to compensate the tadpole terms,
\bea
\label{eq:tadpole1}
& & V_{D6/O6} = - \, 2\, \, e^{4D}\, \left[{\rm Im}N^K \right] \, \Sigma_K \, ,
\eea
where three-form $\Sigma = \Sigma_K \, \beta^K$ is defined through \cite{Aldazabal:2006up, Blumenhagen:2015lta},
\bea
& & \Sigma \equiv {\cal D} F_{RR} = \left(H_K \, m_{0} - \, \omega_{a K}\, m^a + \, Q^a{}_K \, e_a - \, R_K \, e_0 \right)\, \beta^K \,.
\eea
Note that here we have used the expansion of the NS-NS three form $H_3$ and the various RR forms as in eqn. (\ref{eq:flux-components}) and also the flux actions on harmonic forms as given in eqn. (\ref{eq:fluxActions0}). In principle, there can be many other tadpole contributions and exotic branes in the presence of the so-called $P$-fluxes as have been studied in \cite{Aldazabal:2010ef, Aldazabal:2011yz, Lombardo:2016swq, Lombardo:2017yme}.

Now considering the relevant flux actions for a concrete type IIA orientifold setup as given in eqn. (\ref{eq:fluxActions0}), ensuring the nilpotency of the twisted differential ${\cal D}$ on the harmonic forms (via ${\cal D}^2 = 0$) results in the following Bianchi identities \cite{Robbins:2007yv},
\bea
\label{eq:bianchids2}
& & H_K\, \hat{\omega}_\alpha{}^K = 0, \quad H_K\, \hat{Q}^{\alpha K} = 0, \quad \omega_{aK}\, \hat{\omega}_\alpha{}^K = 0, \quad \omega_{aK}\, \hat{Q}^{\alpha K} =0, \nonumber\\
& & R_K\, \hat{Q}^{\alpha K} = 0, \quad R_K \, \hat{\omega}_\alpha{}^K = 0, \quad Q^a{}_K \, \hat{Q}^{\alpha K} = 0, \quad \hat{\omega}_\alpha{}^{K} \,Q^a{}_K=0,\\
& & \left(\hat{d}^{-1}\right)_\alpha{}^\beta\, \, \, \hat{\omega}_\beta{}^{[K}\, \hat{Q}^{\alpha J]} = 0, \quad f^{-1}\, H_{[K} \, R_{J]} - \left(d^{-1}\right)_a{}^b\, \, \, \omega_{b[K}\, Q^a{}_{J]} = 0\,. \nonumber
\eea
Note that in our current conventions, the flux components $H^K$, $R^K$, $\omega_a{}^K, Q^{aK}, \hat{\omega}_{\alpha K}$ and $\hat{Q}^\alpha{}_K$ are projected out, and so they do not appear in the above Bianchi identities. 


\section{New generalized flux orbits and the scalar potential}
\label{sec_NeworbitsAndscalarpotential}
\subsection{Generalized flux orbits}
Using the definitions of the chiral variables given in eqn. (\ref{eq:chiralvariables}), let us consider the RR-flux generated superpotential $W_{RR}$ in eqn. (\ref{eq:Wgen}) which takes the following form,
\bea
\label{eq:Wflux-orbitsRR}
& & W_{RR} = \biggl[\left({\mathbb F}_0 - \sigma_a \, {\mathbb F}^{a} \right) + i \, \left( {\mathbb F}_{a} \, t^a - {\cal V} \, \, {\mathbb F}^0 \right) \biggr]\,,
\eea
where $\sigma_a = \frac{1}{2} \, \kappa_{abc} \, t^b \, t^c, {\cal V} =  \frac{1}{6} \, \kappa_{abc} \, t^a \,t^b \, t^c$ and we introduce the following flux combinations,
\bea
\label{eq:RRorbits0}
& & {\mathbb F}_0 = f\, e_0 + d_a^b\, b^a\, e_b + \frac{1}{2} \, \kappa_{abc} \, b^a\, b^b \,m^c + \frac{1}{6}\, \kappa_{abc}\, b^a \,b^b\,b^c \, m_0\, , \\
& & {\mathbb F}_a = d_a^b \, e_b + \, \kappa_{abc} \, b^b \,m^c + \frac{1}{2}\, \kappa_{abc}\, b^b\,b^c \, m_0\,, \nonumber\\
& & {\mathbb F}^a = m^a + m_0\, b^a\,, \nonumber\\
& & {\mathbb F}^0 = m_0 \,. \nonumber
\eea
These mixings of various RR flux components corresponding to the $F_0, F_2, F_4$ and $F_6$ field strengths are very well known, e.g. from the effective action computations in \cite{Grimm:2004uq, Benmachiche:2006df}. Such flux combinations have been also utilized to rewrite the scalar potential in the absence of (non-)geometric fluxes in \cite{Carta:2016ynn, Farakos:2017jme}. After considering the NS-NS flux induced superpotential $W_{NS-NS}$ as defined in eqn. (\ref{eq:Wgen}) we can rewrite it as under,
\bea
\label{eq:Wflux-orbitsNS}
& & W_{NS-NS} = 2\, N^K \biggl[\left({\mathbb H}_K - \sigma_a \,  {\mathbb Q}^a{}_K \right) + i \, \left( {\mathbb \mho}_{a K}\, t^a - {\cal V}\, \, {\mathbb R}_K \right) \biggr]\, ,
\eea
where similar to the RR-flux combinations we previously had in eqn. (\ref{eq:RRorbits0}), now we find the following peculiar flux combinations to be relevant,
\bea
\label{eq:NSorbitsNew}
& & {\mathbb H}_K \, \, = H_K + \omega_{a K}\, b^a + \frac{1}{2} \kappa_{abc} b^b \, b^c \,  \left({(d^{-1})}^a{}_d \, Q^d{}_K\right) + \frac{1}{6} \kappa_{abc} \, b^a \, b^b \, b^c \, (f^{-1} \, R_K) \, ,\nonumber\\
& & {\mathbb \mho}_{a K} = \omega_{aK} + \kappa_{abc} b^b \,  \left({(d^{-1})}^c{}_d \, Q^d{}_K\right) + \frac{1}{2} \kappa_{abc} \, b^b\, b^c \, (f^{-1} \, R_K), \nonumber\\
& & {\mathbb Q}^a{}_K = \left({(d^{-1})}^a{}_b \, Q^b{}_K\right)+ \, b^a\, \left(f^{-1} R_K \right) \, , \nonumber\\
& & {\mathbb R}_K \, \,\,= \left(f^{-1}\right) \, R_K \,.
\eea
These are the cohomology version of the flux combinations proposed in \cite{Blumenhagen:2013hva}. Now combining the two observations above and also considering the real/imaginary parts of the $N^K$ moduli, the total superpotential $W$ in eqns. (\ref{eq:Wrrnsns})-(\ref{eq:Wgen}) being generated from the various NS-NS and RR-fluxes takes the following form,
\bea
\label{eq:Wflux-orbits}
& & W = \biggl[{\mathbb G}_0 - \sigma_a \, {\mathbb G}^{a} - 2\, e^{-D}\, {\cal X}^K \, \left( {\mathbb \mho}_{a K}\, t^a - {\cal V}\,\, {\mathbb R}_K \right) \biggr] \\
& & \hskip1cm + \, i \, \biggl[{\mathbb G}_{a} \, t^a - {\cal V} \, \, {\mathbb G}^0 + 2\, e^{-D}\, {\cal X}^K \, \left({\mathbb H}_K - \sigma_a \,  \, {\mathbb Q}^a{}_K \right) \biggr]\,,\nonumber
\eea
where the relevant new flux-combinations are captured to be given as under,
\bea
\label{eq:RRorbitsNew}
& & {\mathbb G}_0 = f\, e_0 + d_a^b\, b^a\, e_b + \frac{1}{2} \, \kappa_{abc} \, b^a\, b^b \,m^c + \frac{1}{6}\, \kappa_{abc}\, b^a \,b^b\,b^c \, m_0\, + \xi^K\, {\mathbb H}_K, \\
& & {\mathbb G}_a = d_a^b \, e_b + \, \kappa_{abc} \, b^b \,m^c + \frac{1}{2}\, \kappa_{abc}\, b^b\,b^c \, m_0\, + \xi^K \, {\mathbb \mho}_{a K}, \nonumber\\
& & {\mathbb G}^a = m^a + m_0\, b^a\, + \xi^K\, {\mathbb Q}^a{}_K \,, \nonumber\\
& & {\mathbb G}^0 = m_0 \, +  \xi^K \, {\mathbb R}_K . \nonumber
\eea
Note that all the previously mentioned purely RR-flux orbits, as given in eqn. (\ref{eq:RRorbits0}), are modified with the presence of generalized NS-NS fluxes, which are the cohomology version of the flux combinations proposed in \cite{Blumenhagen:2013hva}. 

Moreover, let us point out that there are hatted (non-)geometric fluxes $\hat{\omega}_{\alpha}{}^J$ and $\hat{Q}^{\alpha J}$ which could generically also survive under the full orientifold action. However, such fluxes do not appear in the superpotential and can only be present via D-term contributions as seen from eqn. (\ref{eq:DtermGen}). On the same analogy, and also motivated by the results of \cite{Shukla:2015bca, Shukla:2015hpa}, we propose the following useful generalization of these fluxes,
\bea
\label{eq:NSorbitsNewTilde}
& & \hat{\mathbb \mho}_{\alpha}{}^K = \hat\omega_{\alpha}{}^K + \hat\kappa_{a\alpha\beta} \, b^a \,  {(\hat{d}^{-1})}^{\beta}{}_\gamma \, \hat{Q}^{\gamma K}, \\
& & \hat{\mathbb Q}^{\alpha K} = {(\hat{d}^{-1})}^{\alpha}{}_\beta \, \hat{Q}^{\beta K} \, , \nonumber
\eea
where we note that the absent components of various fluxes, which one might have naively anticipated to be present, are projected out; for example, $R^K$ does not appear in the generalized versions of $\hat{\mho}_{\alpha}{}^K$ and $\hat{Q}^{\alpha K}$. We will discuss about these more in one of our explicit examples later on.

\subsubsection*{Bianchi identities using new generalized flux orbits}
Let us mention that similar to the observation made for type IIB case in \cite{Shukla:2016xdy}, these generalized NS-NS flux combinations given in eqns. (\ref{eq:NSorbitsNew}) and (\ref{eq:NSorbitsNewTilde}) also respect the set of Bianchi identities given in eqn. (\ref{eq:bianchids2}), which can be expressed as under,
\bea
\label{eq:bianchids3}
& & {\mathbb H}_K\, \hat{{\mathbb \mho}}_\alpha{}^K = 0, \quad {\mathbb H}_K\, \hat{{\mathbb Q}}^{\alpha K} = 0, \quad {\mathbb \mho}_{aK}\, \hat{{\mathbb \mho}}_\alpha{}^K = 0, \quad {\mathbb \mho}_{aK}\, \hat{{\mathbb Q}}^{\alpha K} =0, \nonumber\\
& & {\mathbb R}_K\, \hat{{\mathbb Q}}^{\alpha K} = 0, \quad {\mathbb R}_K \, \hat{{\mathbb \mho}}_\alpha{}^K = 0, \quad {\mathbb Q}^a{}_K \, \hat{{\mathbb Q}}^{\alpha K} = 0, \quad \hat{{\mathbb \mho}}_\alpha{}^{K} \,{\mathbb Q}^a{}_K=0,\\
& & \hat{{\mathbb \mho}}_\alpha{}^{[K}\, \hat{{\mathbb Q}}^{\alpha J]} = 0, \quad {\mathbb H}_{[K} \, {\mathbb R}_{J]} -\, {\mathbb \mho}_{a[K}\, {\mathbb Q}^a{}_{J]} = 0\,. \nonumber
\eea
This happens to be true because of the fact that there exist a generalized version of the twisted differential operator ${\cal D}$ defined in eqn. (\ref{eq:twistedD}) which now reads as \cite{Blumenhagen:2013hva, Blumenhagen:2015lta},
\bea
\label{eq:twistedDnew}
& & {\mathfrak D} = d + {\mathbb H} \wedge . - {\mathbb \mho} \triangleleft . + {\mathbb Q} \triangleright . - {\mathbb R} \bullet .
\eea
Therefore all the above identities follows from the nilpotency of ${\mathfrak D}$ instead of the nilpotency of ${\cal D}$. Let us also note that the three-form combination $\Sigma_K$ appearing in the tadpole contribution given in eqn. (\ref{eq:tadpole1}) remains unchanged after considering the following new generalized version of $\Sigma_K$,
\bea
\label{eq:NewSigma}
& & {\bf \Sigma} \equiv {\bf \Sigma}_K \beta^K = \left({\mathbb H}_K \, {\mathbb G}^{0} - \, {\mathbb \mho}_{a K}\, {\mathbb G}^a + \, {\mathbb Q}^a{}_K \, {\mathbb G}_a - \, {\mathbb R}_K \, {\mathbb G}_0 \right) \, \beta^K \,,
\eea
up to satisfying a subset of these NS-NS Bianchi identities. All the new generalized fluxes mentioned here are given in eqns. (\ref{eq:NSorbitsNew}) and (\ref{eq:RRorbitsNew}).

\subsection{A useful rearrangement of the scalar potential}
In this section we will show that the new generalized flux combinations defined in eqns. (\ref{eq:RRorbits0}), (\ref{eq:NSorbitsNew}), (\ref{eq:RRorbitsNew}) and (\ref{eq:NSorbitsNewTilde}) crucially help in formulating the general scalar potential. A straight forward but tedious computation leads to the reshuffling of the $F$-term scalar potential in the following manner,
\bea
\label{eq:VF_reshuffle}
& & V_{F} = V_{{\mathbb H} {\mathbb H}} + V_{{\mathbb \mho} {\mathbb \mho}} + V_{{\mathbb Q} {\mathbb Q}} + V_{{\mathbb R} {\mathbb R}}  + V_{{\mathbb H} {\mathbb Q}} + V_{{\mathbb R} {\mathbb \mho}} \\
& & \hskip2cm + V_{{\mathbb G}^0 {\mathbb G}^0} + V_{{\mathbb G}^a {\mathbb G}^a} + V_{{\mathbb G}_a {\mathbb G}_a} + V_{{\mathbb G}_0 {\mathbb G}_0} + V_{D6/O6} \, , \nonumber
\eea
where the first line involves the generalized NS-NS flux combinations defined in (\ref{eq:NSorbitsNew}) while the second line, in addition also involves new generalized RR flux combinations given in eqn. (\ref{eq:RRorbitsNew}). The explicit expressions for these pieces are given as under,
\bea
\label{eq:VF_reshuffleNSNS}
& & V_{{\mathbb H} {\mathbb H}} = 4\, e^K \, \left[{\mathbb H}_I \, K^{I \ov J} \, {\mathbb H}_J \right] \\
& & V_{{\mathbb \mho} {\mathbb \mho}} = 4\, e^K \, \left[{\mathbb \mho}_{aI} \, K^{I \ov J} \, {\mathbb \mho}_{bJ} \, t^a \, t^b  +  \, e^{-2 D} \, {\mathbb \mho}_{aI} \, {\mathbb \mho}_{bJ} \, {\cal X}^I\, {\cal X}^J \, \left({K}^{a \ov b} - 4 \, t^a t^b \right)\right] \nonumber\\
& & V_{{\mathbb Q} {\mathbb Q}} = 4\, e^K \, \left[{\mathbb Q}^{a}{}_I \, K^{I \ov J} \, {\mathbb Q}^{b}{}_J  \, \sigma_a \, \sigma_b +\, e^{-2 D} \,\,{\mathbb Q}^a{}_I \, {\mathbb Q}^{b}{}_{J} \, {\cal X}^I\, {\cal X}^J \, \left(16\, {\cal V}^2 \, {K}_{a \ov b} - 4\, \sigma_a \, \sigma_b \right)\right] \nonumber\\
& & V_{{\mathbb R} {\mathbb R}} = 4 \, e^K \, \left[{\cal V}^2 \, \, \, {\mathbb R}_I \, K^{I \ov J} \, {\mathbb R}_J \right] \nonumber\\
& & V_{{\mathbb H} {\mathbb Q}} = (-2) \times 4 \, e^K \, \left[ {\mathbb H}_I \, K^{I \ov J} \, {\mathbb Q}^a{}_J \, \sigma_a \, -4 \,\, e^{-2 D} \, {\mathbb H}_I \, {\cal X}^I \, {\cal X}^J \, {\mathbb Q}^a{}_J \, \sigma_a \right] \nonumber\\
& & V_{{\mathbb R} {\mathbb \mho}} = (-2) \times 4 \, e^K \, {\cal V}\, \left[ {\mathbb R}_I \, K^{I \ov J} \, {\mathbb \mho}_{aJ} \, t^a \, -4 \, e^{-2 D} \,\, {\mathbb R}_I \, {\cal X}^I \, {\cal X}^J \, {\mathbb \mho}_{aJ} \, t^a \right] \nonumber
\eea
 and
\bea
\label{eq:VF_reshuffleRR}
& & V_{{\mathbb G}^0 {\mathbb G}^0} = e^K \, \left[ 4\, {\cal V}^2 \, \left({\mathbb G}^0 \right)^2\right], \quad V_{{\mathbb G}^a {\mathbb G}^a} = e^K \, \left[{\mathbb G}^a \, \left(16\, {\cal V}^2 \,{K}_{a \ov b} \right) \, {\mathbb G}^b \right] \, , \\
& & V_{{\mathbb G}_a {\mathbb G}_a}= \, e^K \, \left[{\mathbb G}_a \, {K}^{a \ov b} \, {\mathbb G}_b \right]\, , \quad V_{{\mathbb G}_0 {\mathbb G}_0} = \, e^K \, \left[ 4 \, \left({\mathbb G}_0 \right)^2 \right] \,, \nonumber\\
& & V_{D6/O6} = -\, e^K \left[16 \, {\cal V} \, {\rm Im}(N^K) \, \, {\bf \Sigma}_K \right]  = -\, 2\, e^{4D} \left[{\rm Im}(N^K) \, \, {\bf \Sigma}_K \right] . \nonumber
\eea
Let us mention that in deriving these important and well motivated pieces we have utilized the following useful definitions and relations,
\bea
\label{eq:Kab}
& & \hskip-1.5cm K_{a \ov b} = \frac{\kappa_a \, \kappa_b - 4\, {\cal V} \, \kappa_{ab}}{16\, {\cal V}^2} \, , \qquad  K^{a \ov b} = 2 \, \, t^a \, t^b - 4\, {\cal V} \, \, \kappa^{ab}\, ,
\eea
where the shorthand notations such as $\kappa_a\, t^a = 6\, {\cal V} =  \kappa_{abc} \, t^a \, t^b \, t^c, \, \kappa_{ab} = \kappa_{abc}\, t^c, \, \kappa_{a} = \kappa_{abc} \, t^b \, t^c = 2 \, \sigma_a$ as well as $\kappa^{ab}$ as the inverse of $\kappa_{ab}$, are used whenever needed. 
Notice that in the absence of (non-)geometric fluxes, the collection of scalar potential pieces in eqns. (\ref{eq:VF_reshuffleNSNS}) and (\ref{eq:VF_reshuffleRR}) reduces into the ones presented in \cite{Carta:2016ynn, Farakos:2017jme}.

\subsection{A symplectic formulation of the scalar potential}
Recall that the superpotential given in eqns. (\ref{eq:Wrrnsns})-(\ref{eq:Wgen}), which we have utilized to get the scalar potential pieces, has been purely motivated by the duality arguments without knowing the higher dimensional origin. Let us try to understand the scalar potential pieces from a higher dimensional point of view on the lines of \cite{Blumenhagen:2013hva, Gao:2015nra, Shukla:2015bca, Shukla:2015hpa, Shukla:2016hyy}, i.e. to invoke the higher dimensional kinetic pieces which could give these scalar potential pieces upon dimensional reduction on a generic CY orientifold. For that purpose, we consider the following generalized multi-form potential,
\bea
\label{eq:GRR}
& & \hskip-0.5cm {\mathbb G}_{RR} \equiv {\mathbb G}^{(0)} + {\mathbb G}^{(2)} + {\mathbb G}^{(4)}+ {\mathbb G}^{(6)} = {\mathbb G}^{0} {\bf 1} + {\mathbb G}^a \, \nu_a + {\mathbb G}_a \, \tilde\nu^a + {\mathbb G}_0 \, \Phi_6\, ,
\eea
where the various flux combinations ${\mathbb G}^{0}, {\mathbb G}^{a}, {\mathbb G}_{a}$ and ${\mathbb G}_{0}$ are defined as in eqn. (\ref{eq:RRorbitsNew}). Here we have used the expansion of forms analogous to the ones given in eqn. (\ref{eq:flux-components}). Subsequently, the generalized RR pieces can be rewritten in the following well motivated integral form, 
\bea
\label{eq:intRR}
& & \hskip-1cm  V_{{\mathbb G}^0 {\mathbb G}^0} = \frac{e^{4\phi}}{2 \, {\cal V}^2} \int_{X_3} {\mathbb G}^{(0)} \wedge \ast {\mathbb G}^{(0)}, \qquad  V_{{\mathbb G}^a {\mathbb G}^a} = \frac{e^{4\phi}}{2 \, {\cal V}^2} \int_{X_3} {\mathbb G}^{(2)} \wedge \ast {\mathbb G}^{(2)} \, ,\\
& & \hskip-1cm V_{{\mathbb G}_a {\mathbb G}_a} = \frac{e^{4\phi}}{2 \, {\cal V}^2} \int_{X_3} {\mathbb G}^{(4)} \wedge \ast {\mathbb G}^{(4)}, \qquad  V_{{\mathbb G}_0 {\mathbb G}_0} = \frac{e^{4\phi}}{2 \, {\cal V}^2} \int_{X_3} {\mathbb G}^{(6)} \wedge \ast {\mathbb G}^{(6)} \, , \nonumber
\eea
where for the last term we have used the fact that on a Calabi Yau threefold $\int_{X_3} \Phi_6 \wedge \ast \Phi_6 = {\cal V}^{-1}$. In addition, we have also used $e^K = e^{4\phi}/(8 \, {\cal V}^3)$ and the following integral definitions of ${K}_{a \ov b}$ and ${K}^{a \ov b}$ which have been previously given in eqn. (\ref{eq:Kab}),
\bea
\label{eq:Gab}
& & \hskip-1cm {K}_{a \ov b} = \frac{1}{4{\cal V}} \, \int_{X_3} \nu_a \wedge \ast \nu_b :=  \frac{1}{4{\cal V}} \, {\cal G}_{ab}\, , \quad {K}^{a \ov b} = \, 4 \, {\cal V} \,  \int_{X_3} \, \tilde{\nu}^a \wedge \ast \tilde{\nu}^b :=  {4{\cal V}} \, {\cal G}^{ab} \,. 
\eea
Now, the $V_{D6/O6}$ piece can be written as,
\bea
& & \hskip-1cm V_{D6/O6} = - \, e^K \left[16 \, {\cal V} \, {\rm Im}(N^K) \, \, {\bf \Sigma}_K \right] = - \, 2 \times \frac{e^{4\phi}}{{\cal V}^2} \int_{X_3} \left[{\rm Im}(N^K) \alpha_K \right] \wedge {\mathfrak D} {\mathbb G}_{RR} \,,
\eea
where recall that using eqn. (\ref{eq:GRR}) and the new generalized twisted differential ${\mathfrak D}$ as defined in eqn. (\ref{eq:twistedDnew}), one gets ${\mathfrak D}{\mathbb G}_{RR} = {\bf \Sigma_K} \, \beta^K$, where ${\bf \Sigma_K}$ is defined as in eqn. (\ref{eq:NewSigma}). Now, such a term should arise from a ten-dimensional Chern-Simons term of the following form \cite{Aldazabal:2006up},
\bea
&& S_{cs} \propto \int_{M_4 \times X_{3}} C_7 \wedge  {\mathfrak D} {\mathbb G}_{RR} \,.
\eea
For formulating the analogous generalized NS-NS pieces, we consider the following useful definitions \cite{Grimm:2004uq},
\bea
\label{eq:MIJ}
& & K_{I \ov J} = 2 \, \, e^{2 \, D} \, {\cal M}_{IJ}, \quad \quad \quad K^{I \ov J} = \frac{1}{2} \, e^{-2D}\, {\cal M}^{IJ} \,.
\eea
Note that considering the orientifold actions in eqn. (\ref{eq:OrientifoldOmega}), the relations (\ref{eq:MIJ}) directly follow from the ${\cal N} =2$ integral definitions involving the three-forms as under \cite{Grimm:2004uq},
\bea
\label{eq:mMatrix}
& & \int_{X_3} \beta^I \wedge \ast \, \beta^J = \, {\rm Im}{\cal M}^{IJ} \, , \\
& & \int_{X_3} \alpha_I \wedge \ast \, \beta^J  = \, {\rm Re}{\cal M}_{IK} \, \, {\rm Im}{\cal M}^{KJ} \, , 
\nonumber\\
& & \int_{X_3} \alpha_I \wedge \ast \alpha_J = \, \left( {\rm Im}{\cal M}_{IJ} + {\rm Re}{\cal M}_{IK} \, \, {\rm Im}{\cal M}^{KL} \, \, {\rm Re}{\cal M}_{LJ} \right)\, , \nonumber
\eea
where
\bea
& & {\cal M}_{IJ} = \ov{\cal F}_{IJ} + 2\, i\, \frac{({\rm Im}\,{\cal F})_{IK} \, {\cal X}^K \, ({\rm Im}\,{\cal F})_{JL} \, {\cal X}^L }{{\cal X}^K({\rm Im}\,{\cal F})_{KL} \, {\cal X}^L} \, .
\eea
Recall that in our normalization of the three-forms, ${\cal X}^I$ is a real function of the complex structure moduli while the ${\cal F}_I$'s are pure imaginary functions. This leads to vanishing of ${\rm Re}{\cal M}$ and subsequently there are only two non-vanishing relation in (\ref{eq:mMatrix}) which lead to two relations of eqn. (\ref{eq:MIJ}).

Now using the relations in eqn. (\ref{eq:MIJ}) and these integral forms, the $V_{{\mathbb H}{\mathbb H}}$ and $V_{{\mathbb R}{\mathbb R}}$ pieces can be rewritten as,
\bea
\label{eq:Vnsns}
& & V_{{\mathbb H} {\mathbb H}} = 4\, e^K \left[{\mathbb H}_I \, K^{I \ov J} \, {\mathbb H}_J \right] = \frac{e^{2\phi}}{4 \, {\cal V}^2} \, \left[{\mathbb H}_I \, {\cal M}^{I  J} \, {\mathbb H}_J \right] \\
& & \hskip1cm = \frac{e^{2\phi}}{4 \, {\cal V}^2} \, \left[{\mathbb H}_I \, \left(\int_{X_3} \beta^I \wedge \ast \, \beta^J \right) \, {\mathbb H}_J \right] = \frac{e^{2\phi}}{4 \, {\cal V}^2} \int_{X_3} {\mathbb H}^{(3)} \wedge \ast {\mathbb H}^{(3)} \,, \nonumber
\eea
and
\bea
& & \hskip-0.5cm  V_{{\mathbb R} {\mathbb R}} = 4 \, e^K \, \left[{\cal V}^2 \, \, \, {\mathbb R}_I \, K^{I \ov J} \, {\mathbb R}_J \right] = \frac{e^{2\phi}}{4} \, \left[{\mathbb R}_I \, {\cal M}^{I  J} \, {\mathbb R}_J \right] \\
& & \hskip1cm = \frac{e^{2\phi}}{4} \, \left[{\mathbb R}_I \, \left(\int_{X_3} \beta^I \wedge \ast \, \beta^J \right) \, {\mathbb R}_J \right] = \frac{e^{2\phi}}{4} \int_{X_3} {\mathbb R}^{(3)} \wedge \ast {\mathbb R}^{(3)} \,. \nonumber
\eea
Here apart from assuming that the flux components ${\mathbb H}_K$ and ${\mathbb R}_K$ are constant parameters we have defined ${\mathbb H}^{(3)}$ and ${\mathbb R}^{(3)}$ through ${\mathbb H}^{(3)} = {\mathbb H}_K \, \beta^K$ and ${\mathbb R}^{(3)} = {\mathbb R}_K \, \beta^K = {\mathbb R} \bullet \Phi_6$, which follow from the analogous flux actions given in (\ref{eq:fluxActions0}). Moreover, using the definition (\ref{eq:MIJ}) we can similarly rewrite (at least the first) piece in each of the four type of remaining terms, which are $V_{{\mathbb \mho} {\mathbb \mho}}, V_{{\mathbb Q} {\mathbb Q}}, V_{{\mathbb H} {\mathbb Q}}$ and $V_{{\mathbb R} {\mathbb \mho}}$. This leads to the following reshuffling,
\bea
\label{eq:partialVint}
& & V_{{\mathbb \mho} {\mathbb \mho}} = \frac{e^{2\phi}}{4 \, {\cal V}^2} \, \int_{X_3} {\mathbb \mho}^{(3)} \wedge \ast {\mathbb \mho}^{(3)}  + \frac{e^{2\phi}}{2 \, {\cal V}^2} \biggl[{\mathbb \mho}_{aI} \, {\mathbb \mho}_{bJ} \, {\cal X}^I\, {\cal X}^J \, \left({K}^{a \ov b} - 4 \, t^a t^b \right) \biggr] \, , \\
& & V_{{\mathbb Q} {\mathbb Q}} =  \frac{e^{2\phi}}{4 \, {\cal V}^2} \, \int_{X_3} {\mathbb Q}^{(3)} \wedge \ast {\mathbb Q}^{(3)}  + \frac{e^{2\phi}}{2 \, {\cal V}^2} \biggl[{\mathbb Q}^a{}_I \, {\mathbb Q}^{b}{}_{J} \, {\cal X}^I\, {\cal X}^J \, \left(16\, {\cal V}^2 \, {K}_{a \ov b} - 4\, \sigma_a \, \sigma_b \right) \biggr] \, , \nonumber\\
& & V_{{\mathbb H} {\mathbb Q}} = \frac{e^{2\phi}}{4 \, {\cal V}^2} \, \int_{X_3} \left(-2\right) \times \, {\mathbb H}^{(3)} \wedge \ast {\mathbb Q}^{(3)}  + \frac{e^{2\phi}}{2 \, {\cal V}^2} \biggl[8 \, \, {\mathbb H}_I \, {\cal X}^I \, {\cal X}^J \, {\mathbb Q}^a{}_J \, \sigma_a \biggr] \, , \nonumber\\
& & V_{{\mathbb R} {\mathbb \mho}} =  \frac{e^{2\phi}}{4 \, {\cal V}^2} \, \int_{X_3} \left(-2 \, {\cal V}\right) \times \, {\mathbb R}^{(3)} \wedge \ast {\mathbb \mho}^{(3)}  + \frac{e^{2\phi}}{2 \, {\cal V}^2} \biggl[8 \, \, {\cal V} \, \, {\mathbb R}_I \, {\cal X}^I \, {\cal X}^J \, {\mathbb \mho}_{aJ} \, t^a \biggr] \, . \nonumber
\eea
Here we have used the shorthand notations ${\mathbb \mho}^{(3)} = {\mathbb \mho}_{a J}\, t^a  \, \beta^J$ and ${\mathbb Q}^{(3)} = {\mathbb Q}^a{}_J  \,\sigma_a\, \beta^J$. Now rewriting the remaining pieces in the integral form is not as easy as it has been for the other pieces so far. Motivated by the type IIB studies in \cite{Shukla:2015hpa, Shukla:2016hyy}, now we utilize the following useful relations\footnote{The analogous identities proposed and proven in \cite{Shukla:2015hpa, Shukla:2016hyy} have a factor of $e^{K_{cs}}$. However, in our current setting we have normalized the three-form $\Omega$ as $\int_{X_3} \, i \, \Omega \wedge \ov\Omega = 1$, and therefore identities in eqn. (\ref{eq:SMid1}) do not have any such factor. We have cross-checked and verified these identities for the two concrete examples to be discussed later on in the upcoming sections.},
\bea
\label{eq:SMid1}
& & \hskip-1.2cm \left({\cal X}^I \, {\cal X}^J \right) = \frac{1}{4}\, {\cal S}^I{}_K \, {\cal M}^{KJ} , \qquad \left({\cal X}^I \, {\cal X}^J \right) = -\, \frac{1}{8} \, {\cal S}^I{}_K \, {\cal M}^{KL} \, {\cal S}_L{}^J \, ,
\eea
where we define the various ${\cal S}$-matrices through the following relations,
\bea
\label{eq:SmatrixDef}
& & \hskip-0.0cm {\cal S}^{IJ} = 4\,{\rm Im}\left({\cal X}^I \ov{\cal X}^J\right), \quad {\cal S}^{I}{}_{J} = 4\,{\rm Im}\left({\cal X}^I \ov{{\cal F}}_J\right), \\
& & \quad {\cal S}_{I}{}^{J} = 4\, {\rm Im}\left({\cal F}_I \ov{\cal X}^J\right) = - \, \left({\cal S}^J{}_I\right)^T, \quad {\cal S}_{IJ} = 4\,{\rm Im}\left({\cal F}_I \ov{{\cal F}}_J\right) \,. \nonumber
\eea
Note that in our construction we have ${\cal X}^I$ as real functions of complex structure moduli while ${\cal F}_I$'s are pure imaginary. Therefore, for our present case, we have
\bea
& & {\cal S}^{IJ} = 0, \quad {\cal S}_{IJ} = 0, \quad {\rm Im}\left({\cal X}^I \, \ov{\cal X}^J \right) = 0, \quad {\rm Im}\left({\cal F}_I \, \ov{\cal F}_J \right) = 0 \,.
\eea
Now using these crucial relations we finally managed to formulate the remaining pieces in the following manner,
\bea
\label{eq:partialVint1}
& & \hskip-1cm V_{{\mathbb \mho} {\mathbb \mho}} = \frac{e^{2\phi}}{4 \, {\cal V}^2} \, \int_{X_3} {\mathbb \mho}^{(3)} \wedge \ast {\mathbb \mho}^{(3)} + \frac{e^{2\phi}}{4 \, {\cal V}} \int_{X_3} \, {\cal G}^{ab} \, \tilde{\mathbb \mho}_a \, \wedge \ast \tilde{\mathbb \mho}_b  - \frac{e^{2\phi}}{4 \, {\cal V}^2} \, \int_{X_3} \tilde{\mathbb \mho}^{(3)} \wedge \ast \tilde{\mathbb \mho}^{(3)} \, , \\
& & \hskip-1cm V_{{\mathbb Q} {\mathbb Q}} = \frac{e^{2\phi}}{4 \, {\cal V}^2} \, \int_{X_3} {\mathbb Q}^{(3)} \wedge \ast {\mathbb Q}^{(3)} + \frac{e^{2\phi}}{4 \, {\cal V}} \int_{X_3} \, {\cal G}_{ab} \, \tilde{\mathbb Q}^a \, \wedge \ast \tilde{\mathbb Q}^b  - \frac{e^{2\phi}}{4 \, {\cal V}^2} \, \int_{X_3} \tilde{\mathbb Q}^{(3)} \wedge \ast \tilde{\mathbb Q}^{(3)} \, , \nonumber\\
& & \hskip-1cm V_{{\mathbb H} {\mathbb Q}} = \frac{e^{2\phi}}{4 \, {\cal V}^2} \, \int_{X_3} \left(-2\right) \, {\mathbb H}^{(3)} \wedge \ast {\mathbb Q}^{(3)} + \frac{e^{2\phi}}{4 \, {\cal V}^2} \, \int_{X_3} \, (- 4)  \, \, {\mathbb H}^{(3)} \wedge \ast \tilde{\mathbb Q}^{(3)} \, , \nonumber\\
& & \hskip-1cm V_{{\mathbb R} {\mathbb \mho}} =  \frac{e^{2\phi}}{4 \, {\cal V}^2} \, \int_{X_3} \left(-2 \, {\cal V}\right) \, {\mathbb R}^{(3)} \wedge \ast {\mathbb \mho}^{(3)} + \frac{e^{2\phi}}{4 \, {\cal V}^2} \, \int_{X_3} \left(- 4 \, {\cal V}\right) \, {\mathbb R}^{(3)} \wedge \ast \tilde{\mathbb \mho}^{(3)}\,, \nonumber
\eea
where we have used the following redefinitions,
\bea
\label{eq:tildefluxes}
\tilde{\mathbb \mho}_a = {\cal S}_K{}^{J} \, {\mathbb \mho}_{a J}  \, \beta^K \,, \quad \tilde{\mathbb Q}^a = {\cal S}_K{}^{J} \, {\mathbb Q}^a{}_J  \, \beta^K\, ; \quad \tilde{\mathbb \mho}^{(3)} = \tilde{\mathbb \mho}_a\, t^a\, , \quad \tilde{\mathbb Q}^{(3)} = \tilde{\mathbb Q}^a \sigma_a.
\eea 
Now we come to look for the integral form of the $D$-terms given in eqn. (\ref{eq:DtermGen}) which can be rewritten as,
\bea
\label{eq:DtermGenNew}
& & \hskip-1.5cm V_D = - 2\, e^{2\, D} \, {\cal F}_J \, {\cal F}_K \, \biggl[\left({\rm Re}(f_g^{\rm ele})^{\alpha\beta} \, \hat{\mho}_\alpha{}^K \, \hat{\mho}_\beta{}^J \right) + \left({\rm Re}{(f_g^{\rm mag})}_{\alpha\beta} \, \hat{\mathbb Q}^{\alpha K} \, \hat{\mathbb Q}^{\beta J} \right)\biggr] \,.
\eea
Similar to the odd-sector moduli space matrix ${\cal G}_{a b}$ (and its inverse ${\cal G}^{a b}$) as given in eqn. (\ref{eq:Gab}), we can analogously define the orientifold even sector moduli-space matrix ${\cal G}_{\alpha\beta}$ (and its inverse matrix ${\cal G}^{\alpha\beta}$) following from the truncation in the K\"ahler metric of the ${\cal N} =2$ sector \cite{Grimm:2004ua, Grimm:2004uq}. These are given as under,
\bea
\label{eq:Galphabeta}
&  {K}_{\alpha \ov \beta} = & \frac{1}{4{\cal V}} \, \int_{X_3} \mu_\alpha \wedge \ast \mu_\beta :=  \frac{{\cal G}_{\alpha\beta}}{4{\cal V}}\, , \\
&  {K}^{\alpha \ov \beta} = & 4 \, {\cal V} \,  \int_{X_3} \, \tilde{\mu}^\alpha \wedge \ast \tilde{\mu}^\beta :=  {4{\cal V}} \, {\cal G}^{\alpha\beta} \, , \nonumber
\eea
where
\bea
& & K_{\alpha \ov{\beta}} = -\frac{3}{2} \frac{\kappa_{\alpha\beta}}{\kappa}= - \, \frac{\kappa_{\alpha\beta}}{4 \, {\cal V}}, \qquad K^{\alpha \ov{\beta}} = - \, 4\, {\cal V} \, \hat\kappa^{\alpha\beta} \,.
\eea
Now, the gauge couplings can be written using the moduli space matrix ${\cal G}_{\alpha\beta}$ and its inverse ${\cal G}^{\alpha\beta}$ as under,
\bea
& & \hskip-1cm {\rm Re}(f_g^{\rm ele})^{\alpha\beta} = - \left(\hat{\kappa}_{a\alpha\beta}\, t^a \right)^{-1} = {\cal G}^{\alpha\beta}, \qquad {\rm Re}(f_g^{\rm mag})_{\alpha\beta} = - {\hat\kappa}_{a\alpha\beta} \, t^a= {\cal G}_{\alpha\beta} \,.
\eea
Note that ${\cal F}_K$'s are pure imaginary and moduli space metrics being positive semi-definite, implies that $V_D \geq 0$. Similar to eqn. (\ref{eq:SMid1}), we have the following identity for $\left({\cal F}_I \, {\cal F}_J \right)$,
\bea
\label{eq:SMid2}
& & \hskip-1.2cm \left({\cal F}_I \, {\cal F}_J \right) = \, \frac{1}{4}\, {\cal S}_I{}^K \, {\cal M}_{KJ} , \qquad \left({\cal F}_I \, {\cal F}_J \right) =  \, \frac{1}{8} \, {\cal S}_I{}^K \, {\cal M}_{KL} \, {\cal S}^L{}_J \,.
\eea
Now using the second identity leads to the following version of $V_D$,
\bea
\label{eq:DtermInt}
& &  V_D = \, \frac{e^{2\phi}}{4\, {\cal V}^2} \, \biggl[ {\cal V} \, {\cal G}^{\alpha\beta}\, \tilde{\hat{\mho}}_\alpha  \wedge \ast\, \tilde{\hat{\mho}}_\beta + {\cal V} \,{\cal G}_{\alpha\beta} \, \tilde{\hat{\mathbb Q}}^{\alpha} \wedge \ast  \tilde{\hat{\mathbb Q}}^{\beta} \biggr] \,,
\eea
where as before in eqn.(\ref{eq:tildefluxes}), now we have defined $\tilde{\hat{\mathbb \mho}}_\alpha = {\cal S}^K{}_{J} \, {\hat{\mathbb \mho}}_\alpha{}^J  \, \alpha_K \,, \, \, \tilde{\hat{{\mathbb Q}}}^\alpha = {\cal S}^K{}_{J} \, \hat{\mathbb Q}^{\alpha J}  \, \alpha_K$.

\subsubsection*{Summary}
We present the following symplectic formulation of the total scalar potential,
\bea
\label{eq:VscalarIntegral}
& & \hskip-1.0cm V_{tot} \equiv V_F + V_D \\
& & \nonumber\\
& &  \hskip-0.2cm = \frac{e^{4\phi}}{2 \, {\cal V}^2} \int_{X_3} \biggl[{\mathbb G}^{(0)} \wedge \ast {\mathbb G}^{(0)} + {\mathbb G}^{(2)} \wedge \ast {\mathbb G}^{(2)} + {\mathbb G}^{(4)} \wedge \ast {\mathbb G}^{(4)} + {\mathbb G}^{(6)} \wedge \ast {\mathbb G}^{(6)} \biggr] \nonumber\\
& & + \frac{e^{2\phi}}{4 \, {\cal V}^2} \int_{X_3} \,\biggl[{\mathbb H}^{(3)} \wedge \ast {\mathbb H}^{(3)} +  {\mathbb \mho}^{(3)} \wedge \ast {\mathbb \mho}^{(3)} + {\mathbb Q}^{(3)} \wedge \ast {\mathbb Q}^{(3)} + {\cal V}^2 \, {\mathbb R}^{(3)} \wedge \ast {\mathbb R}^{(3)} \nonumber\\
& & -2 \, {\mathbb H}^{(3)} \wedge \ast {\mathbb Q}^{(3)} -2 \, {\cal V} \, {\mathbb R}^{(3)} \wedge \ast {\mathbb \mho}^{(3)}  + \, {\cal V} \, {\cal G}^{ab} \, \tilde{\mathbb \mho}_a \, \wedge \ast \tilde{\mathbb \mho}_b + \, {\cal V} \, {\cal G}_{ab} \, \tilde{\mathbb Q}^a \, \wedge \ast \tilde{\mathbb Q}^b \nonumber\\
& & + \, {\cal V} \, {\cal G}^{\alpha\beta} \, \tilde{\hat{\mathbb \mho}}_\alpha \, \wedge \ast \tilde{\hat{\mathbb \mho}}_\beta + \, {\cal V} \, {\cal G}_{\alpha\beta} \, \tilde{\hat{\mathbb Q}}^\alpha \, \wedge \ast \tilde{\hat{\mathbb Q}}^\beta - \tilde{\mathbb \mho}^{(3)} \wedge \ast \tilde{\mathbb \mho}^{(3)} - \tilde{\mathbb Q}^{(3)} \wedge \ast \tilde{\mathbb Q}^{(3)} \nonumber\\
& & - \, 4  \, {\mathbb H}^{(3)} \wedge \ast \tilde{\mathbb Q}^{(3)}  - 4 \, {\cal V} \, {\mathbb R}^{(3)} \wedge \ast \tilde{\mathbb \mho}^{(3)}   \biggr] + V_{D6/O6} \,, \nonumber
\eea
where the $V_{D6/O6}$ is the piece needed to cancel the generalized RR tadpoles, and the various forms are expanded as under,
\bea
\label{eq:three-formfluxes}
& & {\mathbb G}^{(0)} = {\mathbb G}^0 \, {\bf 1}, \quad {\mathbb G}^{(2)} = {\mathbb G}^{a} \, \nu_a, \quad {\mathbb G}^{(4)} = {\mathbb G}_a \, \tilde\nu^a, \quad {\mathbb G}^{(6)} = {\mathbb G}_0 \, \Phi_6\, , \\
& & {\mathbb H}^{(3)} = {\mathbb H}_K \, \beta^K, \quad {\mathbb \mho}^{(3)} = ({\mathbb \mho}_{aK}\, t^a) \, \beta^K, \quad {\mathbb Q}^{(3)} = ({\mathbb Q}^a{}_K \sigma_a) \, \beta^K, \quad {\mathbb R}^{(3)} = {\mathbb R}_K \, \beta^K \equiv {\mathbb R}\bullet \Phi_6, \nonumber\\
& & \tilde{\mathbb \mho}_a = {\cal S}_K{}^{J} \, {\mathbb \mho}_{a J}  \, \beta^K \,, \quad \tilde{\mathbb Q}^a = {\cal S}_K{}^{J} \, {\mathbb Q}^a{}_J  \, \beta^K, \quad   \tilde{\hat{\mathbb \mho}}_\alpha = {\cal S}^K{}_{J} \, {\hat{\mathbb \mho}}_\alpha{}^J  \, \alpha_K, \quad \tilde{\hat{{\mathbb Q}}}^\alpha = {\cal S}^K{}_{J} \, \hat{\mathbb Q}^{\alpha J}  \, \alpha_K\, , \nonumber\\
& & \tilde{\mathbb \mho}^{(3)} = \tilde{\mathbb \mho}_a\, t^a\, , \quad \tilde{\mathbb Q}^{(3)} = \tilde{\mathbb Q}^a \sigma_a, \quad {\cal S}^{I}{}_{J} = 4\,{\rm Im}\left({\cal X}^I \ov{{\cal F}}_J\right) = - ({\cal S}_I{}^J)^T \,.\nonumber
\eea
This way of presenting the scalar potential contributions derived from a superpotential clearly reflects the motivation of exploring their higher dimensional version which could lead to this four-dimensional scalar potential after dimensional reduction. This invoking process is what has been called as `dimensional oxidation' \cite{Blumenhagen:2013hva}.

\section{Demonstrating the insights with explicit examples}
\label{sec_twoExs}
In this section we present two concrete toroidal constructions to exemplify the insights of our symplectic formulation, and we will show how the same scalar potential which arises from the (non-)geometric flux superpotential can also be derived from a set of ten-dimensional kinetic pieces as promoted from eqn. (\ref{eq:VscalarIntegral}).

\subsection{Type IIA on a ${\mathbb T}^6/({\mathbb Z}_2 \times {\mathbb Z}_2)$-orientifold}
Let us consider the type IIA compactification on the orientifold of a ${\mathbb T}^6/({\mathbb Z}_2\times {\mathbb Z}_2)$ orbifold, and start by presenting the necessary ingredients for this toroidal orientifold setup \cite{Villadoro:2005cu, Camara:2005dc, Blumenhagen:2013hva}.
\subsubsection*{Fixing the conventions:}
The complexified coordinates on the torus ${\mathbb T}^6$ are defined as 
\bea
& & \hskip-0.5cm z^1 = R^1\, x^1 + i\, R^2\, x^2, \qquad z^2 = R^3\, x^3 + i\, R^4\, x^4, \qquad z^3 = R^5 \, x^5 + i\, R^6 \, x^6 \,,
\eea
where $0 \leq x^i \leq 1$ and $R^i$ denote the circumference of the $i$-th circle. Further, the two ${\mathbb Z}_2$ orbifold actions are defined as:
\bea
& & \theta: \, \left(z^1, \, z^2, \, z^3 \right) \quad \to \quad  \left( -\, z^1, \, -\, z^2, \, z^3 \right)\,, \\
& & \ov\theta: \, \left(z^1, \, z^2, \, z^3 \right) \quad \to \quad  \left( \, z^1, \, - \, z^2, \, - z^3 \right). \nonumber
\eea
In addition an anti-holomorphic involution $\sigma$ is defined by the following action:
\bea
& & \sigma: \, \left(z^1, \, z^2, \, z^3 \right) \quad \to \quad  \left( -\ov z^1, \, -\ov z^2, \, -\ov z^3 \right).
\eea
Note that the six $R^i$'s defining the complex coordinates $z^i$'s determine the three complex structure moduli $u^i$ and three K\"ahler moduli $t^i$ which can be given as,
\bea
\label{eq:TUModelA}
& & \hskip-1cm t^1 = R^1 R^2, \quad t^2 = R^3 R^4, \quad t^3 = R^5 R^6, \quad u^1 = \frac{R^1}{R^2}, \quad u^2 = \frac{R^3}{R^4}, \quad u^3 = \frac{R^5}{R^6}\,.
\eea 
Further, there are three orientifold ``odd" two-forms which can be constructed as,
\bea
& & \hskip-1cm \nu_1 := \frac{i}{2\, R^1\, R^2}\, dz^1 \wedge d\ov z^1 = dx^1 \wedge dx^2\, , \\
& & \hskip-1cm \nu_2 := \frac{i}{2\, R^3\, R^4}\, dz^2 \wedge d\ov z^2 = dx^3 \wedge dx^4\, , \nonumber\\
& & \hskip-1cm \nu_3 := \frac{i}{2\, R^5\, R^6}\, dz^3 \wedge d\ov z^3 = dx^5 \wedge dx^6\, ,\nonumber
\eea
while there are no ``even" two-forms. Moreover, there are three ``even" four-forms given as,
\bea
& & \tilde{\nu}^1 = \nu_2 \wedge \nu_3, \qquad  \tilde{\nu}^2 = \nu_3 \wedge \nu_1, \qquad \tilde{\nu}^3 = \nu_1 \wedge \nu_2\,,
\eea
while no ``odd" four-forms are present in this setup. 
Finally, let us also mention that there are no harmonic one-form and no five-forms in this construction while the ``odd" six-form $\Phi_6$ is given as,
\bea
& & \Phi_6 \equiv \nu_1 \wedge \nu_2 \wedge \nu_3 = dx^1 \wedge dx^2 \wedge dx^3 \wedge dx^4 \wedge dx^5 \wedge dx^6 \,.
\eea
Now the holomorphic $(3, 0)$ form $\Omega$ can be determined by the choice of the coordinates $z^i$'s up to an overall constant factor. The phase is automatically fixed by our choice of anti-holomorphic involution $\sigma$ via $\sigma^\ast (\Omega) = \ov \Omega$ which suggests to consider the following form for the holomorphic three-form $\Omega$,
\bea
& & \Omega = i \, dz^1 \wedge dz^2 \wedge dz^3.
\eea
Comparing the above definition of the three-form ($\Omega$) with the standard relation $\Omega = {\cal X}^K \, \alpha_K - {\cal F}_K\, \beta^K$, we find that the period vector components are given as,
\bea
\label{eq:periodsExA0}
& & \hskip-1.0cm {\cal X}^0 = R^2\, R^4\, R^6, \quad {\cal X}^1 = \, R^2\, R^3\, R^5, \quad {\cal X}^2 = \, R^1\, R^4\, R^5, \quad {\cal X}^3 = \, R^1\, R^3\, R^6, \\
& & \hskip-0.8cm {\cal F}_0 = -\, i \, R^1\, R^3\, R^5, \quad {\cal F}_1 =- i\, R^1\, R^4\, R^6, \quad {\cal F}_2 = - i\, R^2\, R^3\, R^6, \quad {\cal F}_3 =-  i \, R^2\, R^4\, R^5\,. \nonumber
\eea
One can observe that ${\cal X}^K$'s are real while ${\cal F}_K$'s are pure imaginary functions of the complex structure moduli. Also note that here we have defined the orientifold ``even" basis $\alpha_K$ and ``odd" basis $\beta^K$ as under,
\bea
& & \hskip-1.5cm \alpha_0 = 2 \wedge 4 \wedge 6, \quad \alpha_1 = - 2 \wedge 3 \wedge 5, \quad \alpha_2 = - 1 \wedge 4 \wedge 5, \quad \alpha_3 = - 1 \wedge 3 \wedge 6 \,,\\
& & \hskip-1.5cm \beta^0 = 1 \wedge 3 \wedge 5, \quad \beta^1 = - 1 \wedge 4 \wedge 6, \quad \beta^2 = - 2 \wedge 3 \wedge 6, \quad \beta^3 = - 2 \wedge 4 \wedge 5 \,, \nonumber
\eea
where $2\wedge4\wedge6 = dx^2 \wedge dx^4 \wedge dx^6$ etc. As chosen in eqn. (\ref{eq:intersectionBases}) these basis elements are normalized accordingly as $\int_{X_3} \, \alpha_J \, \wedge \beta^K = \delta_J{}^K$. Moreover, the overall scale factor can be further normalized via the second of the two conditions in eqn. (\ref{eq:Omega=1}).  
Subsequently, the normalized period vectors $\left({\cal X}^I, {\cal F}_J\right)$ are given as\footnote{To avoid unnecessarily introducing new notation, we continue to denote the period vectors as $\left({\cal X}^I, {\cal F}_J\right)$ although it was used in eqn. (\ref{eq:periodsExA0}), before the appropriate normalization was taken into account.},
\bea
\label{eq:periodsExA}
& & \hskip-1cm  {\cal X}^0 = \frac{1}{2 \, \sqrt{2}} \, \sqrt{\frac{R^2 \, R^4 \, R^6}{R^1 \, R^3 \, R^5}}, 
\quad {\cal X}^1 = \,\frac{1}{2 \, \sqrt{2}} \, \sqrt{\frac{R^2 \, R^3 \, R^5}{R^1 \, R^4 \, R^6}} \,,\\ 
& & \hskip-1cm {\cal X}^2 = \, \frac{1}{2 \, \sqrt{2}} \, \sqrt{\frac{R^1 \, R^4 \, R^5}{R^2 \, R^3 \, R^6}}\,,
\quad {\cal X}^3 = \, \frac{1}{2 \, \sqrt{2}} \, \sqrt{\frac{R^1 \, R^3 \, R^6}{R^2 \, R^4 \, R^5}}\,,
\nonumber
\eea
and ${\cal F}_I = -\, i /(8 \, {\cal X}^I)$ for each $I \in \{0, 1, 2, 3\}$, which indeed satisfy ${\cal X}^I \, {\cal F}_I = -i/2$ as could have been expected. Now the complexified variables $N^K$ simplify to the following form,
\bea
& & N^0 = \frac{\xi^0}{2} + \frac{i}{2 \sqrt{2}}\, e^{-\phi} \, R^2\, R^4\, R^6, \qquad N^1 = \frac{\xi^1}{2} + \frac{i}{2 \sqrt{2}}\, e^{-\phi} \, R^2\, R^3\, R^5 \,, \\
& & N^2 = \frac{\xi^2}{2} + \frac{i}{2 \sqrt{2}}\, e^{-\phi} \, R^1\, R^4\, R^5, \qquad N^3 = \frac{\xi^3}{2} + \frac{i}{2 \sqrt{2}}\, e^{-\phi} \, R^1\, R^3\, R^6\,. \nonumber
\eea
This leads to the following expression for the K\"ahler potential,
\bea
& & \hskip-0.5cm K = 4\, D - \, \ln\biggl(-i\, \left(T^1 - \ov T^1\right) \left(T^2 - \ov T^2\right)\left(T^3 - \ov T^3\right) \biggr)\\
& & = -\ln 4 - \sum_{K=0}^3\, \ln \left(N^K - \ov{N}^K \right) \, - \, \ln\biggl(-i\, \left(T^1 - \ov T^1\right) \left(T^2 - \ov T^2\right)\left(T^3 - \ov T^3\right) \biggr) \,. \nonumber
\eea
The superpotential having three $T^a$ moduli and four $N^I$ moduli can be straightly written from the generic expression (\ref{eq:Wgen}) for the choice of $a = \{1, 2, 3 \}$ and $K = \{ 0, 1, 2, 3 \}$. Moreover, let us note that the even $(1,1)$-cohomology is trivial, and subsequently there are no `hatted' fluxes $\hat{\omega}_{\alpha}{}^K, \, \hat{Q}^{\alpha K}$ present in this orientifold setting. Therefore there are no possibility of $D$-term contributions arising from eqn. (\ref{eq:DtermGen}). So the total scalar potential arises from the $F$-term effects subject to satisfying the RR tadpoles constraints and a set of NS-NS Bianchi identities. 

\subsubsection*{A check for our symplectic formulation of the scalar potential}
Let us mention that the $F$-term scalar potential computed from the superpotential results in a total of 2422 terms, and after computing the various pieces using our symplectic formulation we find a perfect match. We present the following useful relations which could be used to directly check or read-off the various scalar potential pieces,
\bea
\label{eq:matricesExA1}
& & {\cal G}^{ab} = {\rm Diag}\left\{\frac{\left(R^1\, R^2\right)^2}{{\cal V}} , \, \frac{\left(R^3\, R^4\right)^2}{{\cal V}}, \, \frac{\left(R^5\, R^6\right)^2}{{\cal V}} \right\} \, , \\
& & {\cal M}^{IJ} = {\rm Diag}\left\{\frac{R^2 R^4 R^6}{R^1 R^3 R^5} , \, \frac{R^2 R^3 R^5}{R^1 R^4 R^6} , \, \frac{R^1 R^4 R^5}{R^2 R^3 R^6} , \,  \frac{R^1 R^3 R^6}{R^2 R^4 R^5} \right\} \, ,
\eea
and 
\bea
\label{eq:matricesExA2}
& & \hskip-1cm {\cal S}^I{}_J = \begin{bmatrix}
\frac{1}{2} &  \quad \frac{R^4 R^6}{2 \, R^3 R^5} & \quad \frac{R^2 R^6}{2 \, R^1 R^5}   &  \quad \frac{R^2 R^4}{2 \, R^1 R^3} \\
& & & \\
\frac{R^3 R^5}{2 \, R^4 R^6} & \quad \frac{1}{2} &  \quad \frac{R^2 R^3}{2 \, R^1 R^4}   & \quad \frac{R^2 R^5}{2 \, R^1 R^6} \\
& & & \\
\frac{R^1 R^5}{2 \, R^2 R^6} & \quad \frac{R^1 R^4}{2 \, R^2 R^3}   & \quad \frac{1}{2} & \quad  \frac{R^4 R^5}{2 \, R^3 R^6} \\
& & & \\
 \frac{R^1 R^3}{2 \, R^2 R^4} & \quad \frac{R^1 R^6}{2 \, R^2 R^5}  & \quad \frac{R^3 R^6}{2 \, R^4 R^5} & \quad \frac{1}{2} \\
& & & \\
\end{bmatrix}\, , \qquad {\cal S}^I{}_J  = - \, \left({\cal S}_I{}^J \right)^T \,.
\eea
Some more insights could be appreciated by looking at the number of terms in various pieces given in eqn. (\ref{eq:VscalarIntegral}) which can be classified into two types; one is such in which the pieces are cleanly separated, and their number of terms can be enumerated in an isolated fashion as have been presented below,
\bea
\label{eq:countsExA}
& & \hskip-1cm \# \left(V_{{\mathbb G}^0{\mathbb G}^0}\right) = 15, \quad \# \left(V_{{\mathbb G}^a{\mathbb G}^a}\right) = 165, \quad \# \left(V_{{\mathbb G}_a{\mathbb G}_a}\right) = 630, \quad \# \left(V_{{\mathbb G}_0{\mathbb G}_0}\right) = 820, \nonumber\\
& & \hskip-1cm \# \left(V_{D6/O6}\right) = 128, \quad \# \left(V_{{\mathbb H}{\mathbb H}}\right) = 144, \quad \# \left(V_{{\mathbb R}{\mathbb R}}\right) = 4 \,.
\eea
These number of terms add up into a total of 1906. The second type of pieces have mixed terms causing cancellations across various pieces, and such remaining pieces produce 516 terms,
\bea
& & \# \left(V_{{\mathbb \mho}{\mathbb \mho}} + V_{{\mathbb Q}{\mathbb Q}} + V_{{\mathbb H}{\mathbb Q}}+ V_{{\mathbb R}{\mathbb \mho}}\right) = 516\,.
\eea

\subsection{Type IIA on a ${\mathbb T}^6/{\mathbb Z}_4$-orientifold}
Now let us consider the type IIA compactification on the orientifold of a ${\mathbb T}^6/{\mathbb Z}_4$ orbifold. This type IIA orientifold setup has been considered for a couple of times for different purposes, e.g. regarding (supersymmetric) moduli stabilization in \cite{Ihl:2007ah, Ihl:2006pp}. Here we re-consider this setup as it is useful for demonstrating the D-term effects arising (from DFT-reduction) which were absent in the previous toroidal setup. Unlike the previous model, the even (1, 1)-cohomology is non-trivial in this background, and one has $h^{1,1}_+(X_3/\sigma) = 1$ in the untwisted sector which is an essential ingredient for generating a non-trivial D-term. Let us start by presenting the necessary ingredients for this type IIA orientifold setup.
\subsubsection*{Fixing the conventions:}
 The complexified coordinates on the torus ${\mathbb T}^6$ are defined as 
\bea
& & \hskip-1cm z^1 = x^1 + i\, x^2, \qquad z^2 = x^3 + i\, x^4, \qquad z^3 = x^5 + \left(\frac{1}{2}+ i\, U \right) \, x^6\,,
\eea
where there is a single complex structure modulus $U$. Further, the ${\mathbb Z}_4$ action on the various coordinates is defined as:
\bea
& & \Theta: \, \left(z^1, \, z^2, \, z^3 \right) \quad \to \quad  \left( i\, z^1, \, i \, z^2, \, - z^3 \right)\,.
\eea
In addition an anti-holomorphic involution $\sigma$ is defined by the following action:
\bea
& & \sigma: \, \left(z^1, \, z^2, \, z^3 \right) \quad \to \quad  \left( \ov z^1, \, i \, \ov z^2, \, \ov z^3 \right)\,.
\eea
Let us note here that $\Theta \, . \, \sigma = \sigma \, . \, \Theta^3$ and so the full orientifold action is isomorphic to the dihedral group ${\cal D}_4$ \cite{Blumenhagen:2002gw, Ihl:2006pp}. In the untwisted sector, there are four two-forms $\nu^a$ which are odd under involution $\sigma$,
\bea
& & \nu_1 \, \, := \, \, \frac{i}{2}\, dz^1 \wedge d\ov z^1 = dx^1 \wedge dx^2\, , \\
& & \nu_2 \, \, := \, \, \frac{i}{2}\, dz^2 \wedge d\ov z^2 = dx^3 \wedge dx^4\, , \nonumber\\
& & \nu_3 \, \, := \, \, \frac{i}{2\, U}\, dz^3 \wedge d\ov z^3 = dx^5 \wedge dx^6\, , \nonumber\\
& & \nu_4 \, \, := \, \, \frac{1 - i}{2}\, \left(dz^1 \wedge d\ov z^2 - i \, dz^2 \wedge d\ov z^1 \right) \nonumber\\
& & \hskip0.8cm = \, \, dx^1 \wedge dx^3\,- dx^1 \wedge dx^4\, + dx^2 \wedge dx^3 \, + dx^2 \wedge dx^4, \nonumber
\eea
and there is a single even two-form $\mu_\alpha$ which can be given as,
\bea
& & \mu_1 := \frac{1 + i}{2}\, \left(dz^1 \wedge d\ov z^2 + i \, dz^2 \wedge d\ov z^1 \right) \\
& & \hskip0.7cm = dx^1 \wedge dx^3\,+ dx^1 \wedge dx^4\, - dx^2 \wedge dx^3 \, + dx^2 \wedge dx^4, \nonumber
\eea
In addition, there are four even four-forms $\tilde{\nu}^a$,
\bea
& & \tilde{\nu}^1 = \nu_2 \wedge \nu_3, \qquad \tilde{\nu}^2 = \nu_1 \wedge \nu_3, \qquad \tilde{\nu}^3 = \nu_1 \wedge \nu_2, \qquad \tilde{\nu}^4 = \nu_3 \wedge \nu_4\,,
\eea
and a single odd four-form $\tilde{\mu}^\alpha$ is:
\bea
\tilde{\mu}^1 = \nu_3 \wedge \mu_1 \,.
\eea
The various intersection numbers and the normalization factors for the integral overs forms are given as:
\bea
\label{eq:intExB}
& & \hskip-1cm f = \frac{1}{4}, \qquad \left\{\kappa_{abc}: \quad \kappa_{123} = \frac{1}{4}, \, \, \kappa_{344} =  -1 \right\}, \qquad \left\{\hat{\kappa}_{a\alpha\beta}: \quad \hat{\kappa}_{311} = -1 \right\}, \\
& & \hskip1.5cm d_a{}^b = {\rm Diag} \biggl\{\, \, \frac{1}{4}, \, \, \frac{1}{4}, \, \, \frac{1}{4}, \, \, -1 \biggr\} , \qquad \tilde{d}_\alpha{}^\beta = \left\{ -1 \right\} \, . \nonumber
\eea
Now the orientifold odd six-form is given as:
\bea
& & \Phi_6 = \nu_1 \wedge \nu_2 \wedge \nu_3 = dx^1 \wedge dx^2 \wedge dx^3 \wedge dx^4 \wedge dx^5 \wedge dx^6\,.
\eea
Considering the K\"ahler form $J$, the volume of the Calabi Yau in string frame is given as,
\bea
& & {\cal V} = \frac{1}{3!} \, \int_{X_3} J \wedge J \wedge J = \frac{1}{4} \, t^3 \left( t^1\, t^2 - 2 \,(t^4)^2\right) \,.
\eea
Let us note here that the following K\"ahler cone conditions ensure the positive definiteness of the above volume form,
\bea
& & t^1 > 0\, , \qquad t^2 > 0\, , \qquad t^3 > 0\, , \qquad t^1 \, t^2 - 2 \, (t^4)^2  > 0\,.
\eea
The three-form basis is given as
\bea
& & \hskip-1.0cm \alpha_0 = 1 \wedge 3 \wedge 5 + 1 \wedge 3 \wedge 6+ 1 \wedge 4 \wedge 5 +2 \wedge 3 \wedge 5  - 2 \wedge 4 \wedge 5  -2 \wedge 4 \wedge 6 \,,\\
& & \hskip-1.0cm \alpha_1 = 1 \wedge 3 \wedge 5 + 1 \wedge 4 \wedge 5 + 1 \wedge 4 \wedge 6 +2 \wedge 3 \wedge 5  + 2 \wedge 3 \wedge 6  -2 \wedge 4 \wedge 5 \,,\nonumber\\
& & \hskip-1.0cm \beta^0 = -1 \wedge 3 \wedge 5 + 1 \wedge 4 \wedge 5 + 1 \wedge 4 \wedge 6 +2 \wedge 3 \wedge 5  + 2 \wedge 3 \wedge 6  + 2 \wedge 4 \wedge 5 \,,\nonumber\\
& & \hskip-1.0cm \beta^1 = 1 \wedge 3 \wedge 5 + 1 \wedge 3 \wedge 6 - 1 \wedge 4 \wedge 5  - 2 \wedge 3 \wedge 5  - 2 \wedge 4 \wedge 5  -2 \wedge 4 \wedge 6 \,, \nonumber
\eea
where $1\wedge 3 \wedge 5 = dx^1 \wedge dx^3 \wedge dx^5$ etc., and one can easily check that $\alpha_K$'s are even under involution while $\beta^K$'s are odd under the involution. Now imposing the condition (\ref{eq:Omega=1}), the holomorphic three-form $\Omega$ takes the following form,
\bea
& & \hskip-1cm \Omega = \frac{1 - i}{2 \, \sqrt{U}} \, dz^1 \wedge dz^2 \wedge dz^3 \\
& & \hskip-0.5cm = \frac{1}{2\, \sqrt{U}} \biggl[ \left(\frac{1}{2} + U \right) \, \alpha_0 + \left(\frac{1}{2} - U \right) \, \alpha_1 + i\, \left(\frac{1}{2} + U \right) \, \beta^0 - i \, \left(\frac{1}{2} - U \right) \, \beta^1\biggr] \,, \nonumber
\eea
from which one reads the period vectors to be given as under,
\bea
\label{eq:periodEx2}
& & {\cal X}^0 = \frac{1}{2\, \sqrt{U}} \left(\frac{1}{2} + U \right), \qquad {\cal X}^1 = \frac{1}{2\, \sqrt{U}} \left(\frac{1}{2} - U \right) \, ,\\
& & \hskip-1cm {\cal F}_0 = - \, \frac{i}{2\, \sqrt{U}} \left(\frac{1}{2} + U \right) = -\, i \, {\cal X}^0, \qquad {\cal F}_1 = \frac{i}{2\, \sqrt{U}} \left(\frac{1}{2} - U \right) = \, i \, {\cal X}^1 \,. \nonumber
\eea
One can observe again that the appropriate normalization of the holomorphic three-form $\Omega$ following from eqn. (\ref{eq:Omega=1}) has ensured that ${\cal X}^K$ is real while ${\cal F}_K$ is pure imaginary. The two complexified chiral variables $N^K = \frac{\xi^K}{2} + i\, e^{-D} \, {\cal X}^K$ are given as,
\bea
& & N^0 = \frac{\xi^0}{2} + i\, e^{-D} \, {\cal X}^0, \qquad N^1 = \frac{\xi^1}{2} + i\, e^{-D} \, {\cal X}^1\,.
\eea
Using the period vector as in eqn. (\ref{eq:periodEx2}), we find that $({\cal X}^0)^2 = \frac{1}{2} + ({\cal X}^1)^2$, and therefore the first piece of the K\"ahler potential $K_Q$ is explicitly given as under,
\bea
& & K_{Q} = 4\, D = -2 \ln \biggl[\frac{1}{2} \left\{\left(N^1 - \ov N^1\right)^2 - \left(N^0 - \ov N^0\right)^2 \right\}\biggr] \,.
\eea
Further, the second piece of the K\"ahler potential $K_S$ is given as,
\bea
& & \hskip-0.5cm K_S = - \ln (8 \, {\cal V} ) = \, -\, \ln\left(\frac{4}{3} \, k_{abc}\, t^a t^b t^c \right) \\
& & \hskip0.22cm  = \, - \, \ln\biggl[\frac{i}{4} \left(T^3 - \ov{T}^3 \right) 
\biggl\{\left(T^1 - \ov{T}^1 \right) \left(T^2 - \ov{T}^2 \right) - 2 \, \left(T^4 - \ov{T}^4 \right)^2 \biggr\} \biggr]\, . \nonumber
\eea
The superpotential having four $T^a$ moduli and two $N^I$ moduli can be straightly written from the generic expression (\ref{eq:Wgen}) for the choice of $a = \{1, 2, 3, 4 \}$ and $K = \{ 0, 1 \}$. Moreover, let us note that unlike the previous example, the even $(1,1)$-cohomology is non-trivial, and subsequently there are additional `hatted' fluxes $\hat{\omega}_{\alpha}{}^K, \, \hat{Q}^{\alpha K}$ which are allowed in this orientifold setting and this leads to $D$-term contributions arising from eqn. (\ref{eq:DtermGen}). So the total scalar potential arise from the $F/D$-term effects subject to satisfying the RR tadpoles constraints and a set of NS-NS Bianchi identities. 

\subsubsection*{A check for our symplectic formulation of the scalar potential}
Let us first discuss about the $F$-term scalar potential computed from the superpotential which results in a total of 4174 terms, and after computing the various pieces using our symplectic formulation we find a perfect match. We present the following useful matrices which could be used to read off the various pieces,
\bea
\label{eq:matricesExB1}
& & {\cal G}^{ab} = \frac{1}{{\cal V}} \, \begin{bmatrix}
 {(t^1)}^2  & 2\, {(t^4)}^2 & 0 &  t^1 \, t^4 \\
 2\, {(t^4)}^2  &  {(t^2)}^2  & 0 & t^2 \, t^4 \\
 0 & 0 &  {(t^3)}^2  & 0 \\
 t^1 \, t^4  &  t^2 \, t^4  &  0 & \left(t^1 \, t^2 + 2\, {(t^4)}^2 \right) \\
 \end{bmatrix}\, , \eea
and 
\bea
\label{eq:matricesExB2}
& & \hskip-1cm {\cal M}^{IJ} = \begin{bmatrix}
U + \frac{1}{4\, U} & \qquad \frac{1}{4\, U} - U  \\
\frac{1}{4\, U} - U & \qquad U + \frac{1}{4\, U}  \\
 \end{bmatrix} \, , \qquad {\cal S}^I{}_J = \begin{bmatrix}
 1 + U + \frac{1}{4\, U} & \quad U - \frac{1}{4\, U} \\
 \frac{1}{4\, U} - U & \quad 1 - U - \frac{1}{4\, U}  \\
 \end{bmatrix}\, . 
\eea
Like the previous example, some more insights could be appreciated by looking at the number of terms in various pieces given in eqn. (\ref{eq:VscalarIntegral}), which can be further classified into two types. The first-type consists of a set of pieces which are cleanly separated in such a way that their number of terms are independently enumerated to be as under,
\bea
\label{eq:countsExA}
& & \hskip-1cm \# \left(V_{{\mathbb G}^0{\mathbb G}^0}\right) = 6, \quad \# \left(V_{{\mathbb G}^a{\mathbb G}^a}\right) = 246, \quad \# \left(V_{{\mathbb G}_a{\mathbb G}_a}\right) = 867, \quad \# \left(V_{{\mathbb G}_0{\mathbb G}_0}\right) = 651, \nonumber\\
& & \hskip-1cm \# \left(V_{D6/O6}\right) = 80, \quad \# \left(V_{{\mathbb H}{\mathbb H}}\right) = 588, \quad \# \left(V_{{\mathbb R}{\mathbb R}}\right) = 6\,,
\eea
which counts a total of 2444 terms. The second-type consists of mixed terms causing cancellations across various pieces, and such remaining pieces produce 1730 terms,
\bea
& & \# \left(V_{{\mathbb \mho}{\mathbb \mho}} + V_{{\mathbb Q}{\mathbb Q}} + V_{{\mathbb H}{\mathbb Q}}+ V_{{\mathbb R}{\mathbb \mho}}\right) = 1730\,.
\eea
In addition, we find that there are in total 34 terms arising from the $D$-term contributions,
\bea
& & \# \left(V_{\hat{\mathbb \mho}\hat{\mathbb \mho}}\right) = 26\, , \qquad \# \left(V_{\hat{\mathbb Q}\hat{\mathbb Q}} \right) = 8 \, ,
\eea
which also have a perfect match when computed from the two formulations given in eqn. (\ref{eq:DtermGenNew}) and eqn. (\ref{eq:DtermInt}).

\section{DFT derivation of the scalar potential}
\label{sec_DFT}
Now we plan to provide a higher dimensional evidence of the well motivated structure of the scalar potential pieces given in eqn. (\ref{eq:VscalarIntegral}). In this regard, we will show that such a four-dimensional scalar potential obtained from the GVW flux superpotential can indeed be derived from the dimensional reduction of a higher dimensional theory, namely the Double Field Theory (DFT). For that purpose, we consider the powerful ${\cal N} =2$ results of \cite{Blumenhagen:2015lta} regarding DFT reduction on Calabi Yau threefold, and implement the same in our Type IIA orientifold framework. For more details on the DFT part, we suggest the readers to directly refer to \cite{Blumenhagen:2015lta}, and hereby we simply collect the relevant ingredients needed to establish the connection with our approach. Note that, as we have already converted the total scalar potential into real moduli/axions (and the final collection does not use the complexified fields), so we can directly check the connection by simply considering the orientifold projected version of various terms appearing in the ${\cal N} =2$ DFT Lagrangian compactified on a Calabi Yau threefold. The same (in string-frame) can be given as the sum of following two collection of pieces \cite{Blumenhagen:2015lta},
\bea
S_{10d} \supset \int_{M_4 \times X_3} \, \sqrt{-G} \, \, \left(L_{NS\,\,NS} +  L_{RR} \right) \, ,
\eea
where the two pieces are given as:
\bea
& & \hskip-2cm L_{RR} = -\frac{1}{2} \, {\mathfrak G} \wedge \ast {\mathfrak G} \,, \nonumber\\
& & \hskip-2cm L_{NS \,\,NS} = -\frac{e^{-2 \, \phi}}{4} \biggl[\chi \wedge \ast \ov\chi + \, \Psi \wedge \ast \ov\Psi \\
& & -\frac{1}{2} \Big(\Omega\wedge \chi\Big)\wedge \ast\, \Big(\ov\Omega\wedge \ov \chi\Big)-{1\over 2}\, \Big(\Omega\wedge \ov\chi\Big)\wedge \ast \, \Big(\ov\Omega\wedge  \chi\Big) \biggr] \, . \nonumber
\eea
As we are following the notations of \cite{Grimm:2004ua} in the definitions of moduli space metrics, we have a difference of an overall factor of $1/2$ between the ten-dimensional NS-NS and RR sector kinetic pieces as compared to the proposal of \cite{Blumenhagen:2015lta}. 
Let us first elaborate on the various ingredients by relating them into our conventions:
\begin{itemize} 
\item{The flux combination $\chi$ can be given as,
\bea
\label{eq:chi}
& & \hskip-0.95cm \chi \equiv  {\mathfrak D} \, e^{-J_c} ={\mathbb H} - {\mathbb \mho}\triangleleft (- J_c)+{\mathbb Q}\triangleright
\left({(- J_c)\wedge (- J_c)\over 2}\right) - {\mathbb R}\bullet
     \left({(- J_c)\wedge (- J_c)\wedge (- J_c)\over 6}\right) \nonumber\\
& & \hskip-0.0cm = \biggl[\left({\mathbb H}_K - \sigma_a \, {\mathbb Q}^a{}_K \right) + i\, \left( \mho_{a K} \, t^a - {\cal V} \, \, {\mathbb R}_K \right) \biggr] \, \beta^K \,, 
\eea
where similar to the previously defined twisted differential operator ${\cal D}$, a new operator ${\mathfrak D} = d + \, {\mathbb H} \wedge . - \, {\mathbb \mho} \triangleleft . + \, {\mathbb Q} \triangleright .- \, {\mathbb R} \bullet . $ has been introduced to incorporate the effects of $B_2$-field  such that, 
\bea
& & \hskip-0.5cm {\mathbb H} = {H} + {\omega}\triangleleft B_2 + {Q}\triangleright \left({B_2\wedge B_2\over 2}\right) + {R}\bullet
     \left({B_2\wedge B_2\wedge B_2\over 6}\right) \, \, {\rm etc.}
\eea as have been already presented in cohomology formulation in eqn. (\ref{eq:NSorbitsNew}).
}
\item{In our conventions, the generalized RR three-form field strength ${\mathfrak G}$ is given as,
\bea
\label{eq:mathfrak G}
& & {\mathfrak G}\equiv {\mathbb F} - {\mathfrak D} \, {\cal C} \,,
\eea
where the generalized RR-form potential ${\cal C} = C^{(1)} + C^{(3)}+ C^{(5)} + ... $ and  ${\mathbb F} = {\mathbb F} _{(0)} + {\mathbb F}_{(2)}  + {\mathbb F}_{(4)} + {\mathbb F}_{(6)}$ are the same as defined in eqn. (\ref{eq:RRorbits0}). 
\bea
\label{eq:DFTG}
& & \hskip-0.5cm {\mathfrak G} = {\mathbb F} - {\mathbb H} \wedge C^{(3)} \, + \, \mho \triangleleft C^{(3)} - {\mathbb Q}\triangleright C^{(3)} + {\mathbb R} \bullet C^{(3)} \\
& & = \, {\mathbb G}_0 \, \Phi_6  +  {\mathbb G}^a \, \, \nu_a + {\mathbb G}_a \, \, \tilde{\nu}^a + {\mathbb G}^0 \, {\bf 1} \, , \nonumber
\eea
where in the last line we have utilized $C^{(3)} = \, \xi^K \, \alpha_K$, the flux actions given in eqn. (\ref{eq:fluxActions0}) to get the ${\mathbb G}$ field strengths defined in eqn. (\ref{eq:RRorbitsNew}), and the fact that there are no invariant one-forms and five-forms present in the orientifold construction.}
\item{The third flux combination $\Psi$ is defined as,
\bea
\label{eq:Psi}
& & \hskip-0.5cm \Psi \equiv \, {\mathfrak D}\, \Omega={\mathbb H}\wedge \Omega - {\mathbb \mho}\triangleleft \Omega + {\mathbb Q}\triangleright \Omega - {\mathbb R}\bullet  \Omega \nonumber\\
& & = - \, {\cal X}^K \biggl[f^{-1} \, H_K \, \Phi_6 + \, {(d^{-1})}_a{}^b \, \mho_{b K} \, \, {\tilde\nu}^a + \, {\mathbb Q}^a{}_{K} \, \, {\nu}_a + \, {\mathbb R}_K\, {\bf 1} \biggr]\, \\
& & \hskip1cm - \, i\, {\cal F}_K \biggl[{(\hat{d}^{-1})}_\alpha{}^\beta \, \hat\mho_{\beta}{}^{K} \, \, \tilde{\mu}^\alpha + \, \hat{\mathbb Q}^{\alpha K}\, \, {\mu}_\alpha \biggr] \, , \nonumber
\eea
where in writing the second equality we have used the relation $\Omega = {\cal X}^K\, \alpha_K - {\cal F}^K\, \beta^K$ along with the various flux actions defined in eqn. (\ref{eq:fluxActions0}). However, let us note that we have normalized our three-form $\Omega$ as mentioned in eqn. (\ref{eq:Omega=1}), and therefore while deriving our symplectic formulation from DFT results of \cite{Blumenhagen:2015lta}, we need to take this factor into account. Given that for a Calabi Yau threefold, the following relation holds,
\bea
\label{eq:OmegaNormalization}
\frac{i \, \Omega \wedge \ov \Omega}{8} = \frac{1}{6} \, J \wedge J \wedge J \Longrightarrow \int_{X_3} i\, \Omega \wedge \ov \Omega = 8 {\cal V} \,,
\eea
and therefore in order to satisfy our normalization condition $ \int_{X_3} i\, \Omega \wedge \ov \Omega = 1$, we need to rescale the period vectors $\left({\cal X}^I, {\cal F}_I\right)$ of \cite{Blumenhagen:2015lta} by a factor of $\sqrt{(8 {\cal V})}$. Also note that the imaginary part of $\Psi$ as seen in eqn. (\ref{eq:Psi}) is crucial as it produces the $D$-term contributions. This is because it involves the `hat' index fluxes $\hat\mho_\alpha{}^K$ and $\hat{\mathbb Q}^{\alpha K}$ defined in eqn. (\ref{eq:NSorbitsNewTilde}).}
\end{itemize}
Now,  we will investigate the various terms $L_{NS \,NS}$ and $L_{RR}$ of the DFT reduction to connect with those of ours. 
\subsection{F-term contributions}
\subsubsection*{Matching the generalized RR sector} 
Using the expressions ${\mathfrak G}$ in eqn. (\ref{eq:DFTG}) we find that,
 \bea
 \label{eq:rrVDFT}
 & & \hskip-1.65cm - \, \frac{1}{2} {\mathfrak G} \wedge \ast {\mathfrak G} \equiv - \, \frac{1}{2} \, \, \left({\mathbb G}_0 \right)^2 \Phi_6 \wedge \ast \Phi_6\, - \, \frac{1}{2} \, \, \left({\mathbb G}^a \, {\mathbb G}^b \right) \, \nu_a \wedge \ast \nu_b \\
 & & \hskip1cm - \, \frac{1}{2} \, \, \left({\mathbb G}_a \, {\mathbb G}_b \right) \, \tilde\nu^a \wedge \ast \tilde\nu^b - \, \frac{1}{2} \, \, \left({\mathbb G}^0 \right)^2 \ast {\bf 1}. \nonumber
 \eea
It is easy to observe that the above pieces in eqn. (\ref{eq:rrVDFT}) lead to the 4D scalar potential pieces given in previous eqn. (\ref{eq:intRR}). For confirming the overall factor, we consider the (string frame) ten-dimensional metric $G_{MN} = {\rm BlockDiag}\left[\frac{e^{2\phi}}{{\cal V}} g_{\mu\nu}, \, \, g_{ij} \right]$ and subsequently by taking the integration as under,
\bea
& & \frac{1}{2} \int d^{10}x \sqrt{ - G_{MN}} \left ( ... \right) = \frac{1}{2} \int d^4x \sqrt{g_{\mu\nu}} \, \, \, \frac{e^{4\phi}}{{\cal V}^2} \, \int_{X_3} \left( ... \right) \,.
\eea
Thus considering the overall factor $e^{4\phi}/{(2\, {\cal V}^2)}$, it is straight forward to recover all the pieces present in the first line of eqn. (\ref{eq:VscalarIntegral}) from the ones given in eqn. (\ref{eq:rrVDFT}).

\subsubsection*{Matching the generalized NS-NS sector}
Using the flux combination $\chi$ being defined in eqn. (\ref{eq:chi}), we find the following pieces in string-frame,
\begin{eqnarray}
\label{eq:V11111}
 & & \frac{1}{4} \, {\chi} \wedge \ast \ov{\chi} = \frac{1}{4} \, \, \, \biggl[{\mathbb H}^{(3)} \wedge \ast {\mathbb H}^{(3)} + \, {\mathbb \mho}^{(3)} \wedge {\ast} {\mathbb \mho}^{(3)} + {\mathbb Q}^{(3)} \wedge \ast {\mathbb Q}^{(3)}   + {\cal V}^2\, {\mathbb R}^{(3)} \wedge {\ast} {\mathbb R}^{(3)} \nonumber\\
& & \hskip2.5cm  - 2  \, {\mathbb H}^{(3)} \wedge \ast {\mathbb Q}^{(3)} - 2 \, {\cal V}\, {\mathbb R}^{(3)} \wedge {\ast} {\mathbb \mho}^{(3)} \biggr]\, .
\end{eqnarray}
Note that all these pieces indeed appear explicitly in the integral version of our scalar potential given in eqn. (\ref{eq:VscalarIntegral}). Now considering the multi-degree form $\Psi$ as defined in eqn. (\ref{eq:Psi}), we find that $(\Psi\wedge\ast \ov\Psi)$ have the following pieces,
\begin{eqnarray}
\label{eq:Vk22222}
 & & \hskip-1.5cm \frac{1}{4} \, {\Psi} \wedge \ast \ov{\Psi} = \frac{1}{4} \, \, \, \biggl[{({\mathbb H}\wedge \Omega)} \wedge \ast ({\mathbb H}\wedge \ov\Omega) + ({\mathbb R}\bullet\Omega) \wedge \ast ({\mathbb R}\bullet \ov\Omega)  \\
 & & \hskip2cm + {({\mathbb \mho}\triangleleft \Omega)} \wedge \ast ({\mathbb \mho}\triangleleft \ov\Omega) +({\mathbb Q}\triangleright \Omega) \wedge \ast ({\mathbb Q}\triangleright \ov\Omega)  \biggr] \, . \nonumber
\end{eqnarray}
Further considering the two cross-pieces of $L_{NS\, NS}$ we have,
\begin{eqnarray}
\label{eq:Vk33333}
& & \hskip-0.0cm - \, \frac{1}{8} \bigg[ \Big(\Omega\wedge \chi\Big)\wedge
   *\, \Big(\ov\Omega\wedge \ov \chi\Big)+ \Big(\Omega\wedge \ov\chi\Big)\wedge
   *\, \Big(\ov\Omega\wedge  \chi\Big) \biggr] \nonumber\\
& & \hskip1.3cm = - \, \frac{1}{4}\, \biggl[ \Big(\Omega\wedge {\rm Re}({\chi})\Big)\wedge
   *\, \Big(\ov\Omega\wedge  {\rm Re}({\chi})\Big) + \Big(\Omega\wedge {\rm Im}({\chi})\Big)\wedge
   *\, \Big(\ov\Omega\wedge  {\rm Im}({\chi})\Big) \biggr]\nonumber\\
& & \hskip1.3cm = - \, \frac{1}{4} \biggl[({\mathbb H}^{(3)} \wedge \Omega) \wedge \ast ({\mathbb H}^{(3)} \wedge \ov\Omega) + {\cal V}^2 \, ({\mathbb R}^{(3)} \wedge \Omega) \wedge \ast ({\mathbb R}^{(3)} \wedge \ov\Omega) \nonumber\\
& & \hskip2.5cm + {({\mathbb \mho}^{(3)} \wedge \Omega)} \wedge \ast ({\mathbb \mho}^{(3)} \wedge \ov\Omega) + {({\mathbb Q}^{(3)} \wedge \Omega)} \wedge \ast ({\mathbb Q}^{(3)} \wedge \ov\Omega) \nonumber\\
& & \hskip2.5cm  -{({\mathbb H}^{(3)}\wedge \Omega)} \wedge \ast ({\mathbb Q}^{(3)}\wedge \ov\Omega)  - {({\mathbb Q}^{(3)}\wedge \Omega)} \wedge \ast ({\mathbb H}^{(3)}\wedge \ov\Omega)   \nonumber\\
& & \hskip2.5cm - {\cal V} \, {({\mathbb R}^{(3)} \wedge \Omega)} \wedge \ast ({\mathbb \mho}^{(3)} \wedge \ov\Omega) - {\cal V} \, {({\mathbb \mho}^{(3)} \wedge \Omega)} \wedge \ast ({\mathbb R}^{(3)} \wedge \ov\Omega) \biggr]\,, 
\end{eqnarray}
where in the last equality, we have used the shorthand notations ${\rm Re}(\chi) = {\mathbb H}^{(3)} - {\mathbb Q}^{(3)}$ and ${\rm Im}(\chi) = {\mathbb \mho}^{(3)} - {\cal V} \, {\mathbb R}^{(3)}$ with appropriate indices as given in eqn. (\ref{eq:chi}).  Let us mention that using the flux actions in eqn. (\ref{eq:fluxActions0}), it is easy to see that the two terms in eqn. (\ref{eq:Vk22222}) are canceled by their counter pieces in eqn. (\ref{eq:Vk33333}). Finally, for producing the remaining $F$-term pieces of eqn. (\ref{eq:VscalarIntegral}), i.e. the ones without hatted flux components, from the eqns. (\ref{eq:Vk22222}) and (\ref{eq:Vk33333}), we simply need to consider the following useful relation \cite{Blumenhagen:2015lta},
\bea
\label{eq:wedge-Identity}
& & \hskip-1cm \int_{X_3} \left(\alpha_I \wedge \beta^{I^\prime} \right) \wedge \ast \left(\alpha_J \wedge \beta^{J^\prime} \right) \\
& & = \frac{1}{\cal V} \, \left( \int_{X_3} \alpha_I \wedge \beta^{I^\prime} \right) \left(\int_{X_3} \alpha_J \wedge \beta^{J^\prime} \right) = \frac{\delta_{I}{}^{I^\prime}\, \delta_{J}{}^{J^\prime}}{{\cal V}}\,,\nonumber
\eea
along with taking care of a factor of $\sqrt{(8 {\cal V})}$ in the period vectors $({\cal X}^I, {\cal F}_J)$ as argued in the paragraph below the eqn. (\ref{eq:OmegaNormalization}). Note that the first equality in eqn. (\ref{eq:wedge-Identity}) holds for any generic real six-form on the Calabi Yau threefold, e.g. $\int_{X_3} \Phi_6 \wedge \ast \Phi_6 = {\cal V}^{-1}$.

\subsection{D-term contributions}
The only pieces which we have not covered so far arise from the imaginary part of $\Psi$ flux as given in eqn. (\ref{eq:Psi}). This leads to two additional contributions,
\begin{eqnarray}
& & \hskip-1.0cm \int_{M_4 \times X_3} \frac{e^{-2 \phi}}{4} \left({\rm Im}\Psi \right) \wedge\ast \left({\rm Im}\Psi \right) =  \int_{M_4 \times X_3} \frac{e^{-2 \phi}}{4} \, \times {(8 \, {\cal V})} \times \, {\cal F}_K \, \ov{\cal F}_L \\
& &  \times \biggl[{(\hat{d}^{-1})}_\alpha{}^{\alpha^\prime} \, \hat\mho_{\alpha^\prime}{}^{K} \, \, \tilde{\mu}^\alpha + \, \hat{\mathbb Q}^{\alpha K}\, \, {\mu}_\alpha \biggr] \wedge \ast \biggl[{(\hat{d}^{-1})}_\beta{}^{\beta^\prime} \, \hat\mho_{\beta^\prime}{}^{L} \, \, \tilde{\mu}^\beta + \, \hat{\mathbb Q}^{\beta L}\, \, {\mu}_\beta \biggr] \,, \nonumber
\eea
where we have multiplied a factor of $\sqrt{(8 {\cal V})}$ in the period vectors $({\cal X}^I, {\cal F}_J)$ as argued in the paragraph below the eqn. (\ref{eq:OmegaNormalization}). This induces the following contribution in the four-dimensional scalar potential,
\bea
& & 2\, e^{2D} \, {\cal F}_K \, \ov{\cal F}_L \biggl[\hat\mho_{\alpha}{}^{K} \, {\cal G}^{\alpha\beta} \, \hat\mho_{\beta}{}^{L} \, + \hat{\mathbb Q}^{\alpha K} \, {\cal G}_{\alpha\beta}\,\hat{\mathbb Q}^{\beta L} \biggr] ,\nonumber
\end{eqnarray}
which is indeed the positive definite $D$-term given in eqn. (\ref{eq:DtermGenNew}). This completes the DFT derivation of our symplectic formulation. Thus our current analysis completes the type IIA side of the story initiated in \cite{Blumenhagen:2015lta}.

\section{Conclusions}
\label{sec_conclusions}
In this article we have studied some interesting aspects of the four-dimensional type IIA effective potential in the presence of (non-)geometric fluxes, in addition to the usual NS-NS and RR fluxes. In this regard, first we present some peculiar flux combinations which are given in the eqns. (\ref{eq:NSorbitsNew})-(\ref{eq:NSorbitsNewTilde}). Subsequently, using these flux orbits, we formulate the scalar potential arising from a generic non-geometric flux superpotential in a few pieces. We call this as the `symplectic formulation' of the scalar potential. The main motivation of such a formulation was to invoke the higher dimensional origin of the 4D scalar potential on the lines of \cite{Blumenhagen:2013hva}, which leads to the following form of the scalar potential:
\bea
& & \hskip-1.0cm V_{tot} \equiv  \frac{e^{4\phi}}{2 \, {\cal V}^2} \int_{X_3} \biggl[{\mathbb G}^{(0)} \wedge \ast {\mathbb G}^{(0)} + {\mathbb G}^{(2)} \wedge \ast {\mathbb G}^{(2)} + {\mathbb G}^{(4)} \wedge \ast {\mathbb G}^{(4)} + {\mathbb G}^{(6)} \wedge \ast {\mathbb G}^{(6)} \biggr] \nonumber\\
& & + \frac{e^{2\phi}}{4 \, {\cal V}^2} \int_{X_3} \,\biggl[{\mathbb H}^{(3)} \wedge \ast {\mathbb H}^{(3)} +  {\mathbb \mho}^{(3)} \wedge \ast {\mathbb \mho}^{(3)} + {\mathbb Q}^{(3)} \wedge \ast {\mathbb Q}^{(3)} + {\cal V}^2 \, {\mathbb R}^{(3)} \wedge \ast {\mathbb R}^{(3)} \nonumber\\
& & -2 \, {\mathbb H}^{(3)} \wedge \ast {\mathbb Q}^{(3)} -2 \, {\cal V} \, {\mathbb R}^{(3)} \wedge \ast {\mathbb \mho}^{(3)}  - \, 4  \, {\mathbb H}^{(3)} \wedge \ast \tilde{\mathbb Q}^{(3)}  - 4 \, {\cal V} \, {\mathbb R}^{(3)} \wedge \ast \tilde{\mathbb \mho} \nonumber\\
& &  + \, {\cal V} \, {\cal G}^{ab} \, \tilde{\mathbb \mho}_a \, \wedge \ast \tilde{\mathbb \mho}_b + \, {\cal V} \, {\cal G}_{ab} \, \tilde{\mathbb Q}^a \, \wedge \ast \tilde{\mathbb Q}^b + \, {\cal V} \, {\cal G}^{\alpha\beta} \, \tilde{\hat{\mathbb \mho}}_\alpha \, \wedge \ast \tilde{\hat{\mathbb \mho}}_\beta + \, {\cal V} \, {\cal G}_{\alpha\beta} \, \tilde{\hat{\mathbb Q}}^\alpha \, \wedge \ast \tilde{\hat{\mathbb Q}}^\beta  \nonumber\\
& & - \tilde{\mathbb \mho}^{(3)} \wedge \ast \tilde{\mathbb \mho}^{(3)} - \tilde{\mathbb Q}^{(3)} \wedge \ast \tilde{\mathbb Q}^{(3)}  \biggr] + V_{D6/O6} \,,\nonumber
\eea
where the various form-fluxes are defined as in eqn. (\ref{eq:three-formfluxes}). In order to demonstrate the insights of our symplectic formulation, we have presented the detailed computation for two concrete examples by utilizing the orientifolds of the complex threefolds ${\mathbb T}^6/({\mathbb Z}_2 \times {\mathbb Z}_2)$ and ${\mathbb T}^6/{\mathbb Z}_4$. The first setup, which is very simple and has served as a canonical example in many previous studies pertaining to various different aspects, illustrates many explicit ingredients of our proposal. However it fails to incorporate the $D$-terms contributions because of the fact that the even sector for the $(1,1)$-cohomology is trivial and therefore all the fluxes which could contributed in the $D$-term scalar potential are projected out. However, the second example is rich enough to have a non-trivial even sector for the $(1,1)$-cohomology with $h^{1,1}_+(X_3/\sigma) = 1$ in the untwisted sector, and hence does produce a non-trivial $D$-term contribution in support of our formulation. 

Secondly, we have also shown how our symplectic formulation can be connected to the robust ${\cal N} =2$ results of the Double Field Theory reduction on Calabi Yau threefolds \cite{Blumenhagen:2015lta}. A similar symplectic analysis for the type IIB case, and its DFT connections have been also presented in \cite{Shukla:2016hyy}.  In this regard, we would like to make the following observations:
\begin{itemize}
\item{The various pieces of the scalar potential collected in our symplectic formulation can indeed be derived from a higher dimensional theory (namely DFT) after compactifying the same on a Calabi Yau orientifold.}
\item{Unlike the type IIB case in which the $D$-terms arise from the imaginary part of the flux combination $\chi$ \cite{Blumenhagen:2015lta, Shukla:2015hpa}, we find that in the current type IIA setting, the $D$-terms arise from the imaginary part of the flux combination $\Psi$.}
\item{Our analysis provides the type IIA counterpart regarding the checks for the robust ${\cal N} =2$ results obtained from the DFT reduction on Calabi Yau threefolds \cite{Blumenhagen:2015lta}.} 
\end{itemize} 
To summarize, it is quite interesting to note that the same scalar potential can be derived from three different routes; (i) directly from the flux superpotential, (ii) from our symplectic formulation and (iii) from the DFT reduction on Calabi Yau orientifolds. We have illustrated these routes in two concrete examples.
Using the compact form of the scalar potential presented in our symplectic formulation, it would be possible to perform a model independent study of non-supersymmetric moduli stabilization in the generic non-geometric type IIA orientifold setups, and we leave this for the future work.

\section*{Acknowledgments}
We thank Michael Fuchs, Fernando Marchesano and Wieland Staessens for several useful discussions. We also gratefully thank Ralph Blumenhagen for his encouraging comments on this work. The work of PS has been supported by the ERC Advanced Grant ``String Phenomenology in the LHC Era" (SPLE) under contract ERC-2012-ADG-20120216-320421.



\bibliographystyle{utphys}
\bibliography{reference}

\end{document}